\documentclass[sn-mathphys-num]{sn-jnl}% Math and Physical Sciences Numbered Reference Style 
\usepackage{graphicx}%
\usepackage{multirow}%
\usepackage{amsmath,amssymb,amsfonts}%
\usepackage{amsthm}%
\usepackage{mathrsfs}%
\usepackage[title]{appendix}%
\usepackage{xcolor}%
\usepackage{textcomp}%
\usepackage{manyfoot}%
\usepackage{booktabs}
\usepackage{algorithm}%
\usepackage{algorithmicx}%
\usepackage{algpseudocode}%
\usepackage{listings}%
\usepackage{adjustbox}
\definecolor{milkgreen}{rgb}{0.5,1,0.5}
\usepackage{array}
\usepackage{fullpage}
\usepackage[most]{tcolorbox}
\usepackage{wrapfig}
\usepackage{enumitem}
\usepackage{boldline}
\usepackage{colortbl}
\usepackage{makecell}
\usepackage{caption}

\theoremstyle{thmstyleone}%
\theoremstyle{thmstyletwo}%

\theoremstyle{thmstylethree}%

\begin{document}

\title[Digital Forensics in the Age of Large Language Models]{Digital Forensics in the Age of Large Language Models}
% \title[Large Language Models for Digital Forensics]{Large Language Models for Digital Forensics}

%\author{Anonymous author}

\author*[1]{\fnm{Zhipeng} \sur{Yin}}\email{zyin007@fiu.edu}

\author[1]{\fnm{Zichong} \sur{Wang}}

\author[2]{\fnm{Weifeng} \sur{Xu}}
\author[3]{\fnm{Jun} \sur{Zhuang}}
\author[1]{\fnm{Pallab} \sur{Mozumder}}
\author[1]{\fnm{Antoinette} \sur{Smith}}
\author[1]{\fnm{Wenbin} \sur{Zhang}}%\email{wenbin.zhang@fiu.edu}

\affil[1]{\orgname{Florida International University}, \orgaddress{\city{Miami},  \state{Florida}, \country{USA}}}
\affil[2]{\orgname{University of Baltimore}, \orgaddress{\city{Baltimore},  \state{Maryland}, \country{USA}}}
\affil[3]{\orgname{Boise State University}, \orgaddress{\city{Boise}, \state{Idaho}, \country{USA}}}


\abstract{Digital forensics plays a pivotal role in modern investigative processes, utilizing specialized methods to systematically collect, analyze, and interpret digital evidence for judicial proceedings. However, traditional digital forensic techniques are primarily based on manual labor-intensive processes, which become increasingly insufficient with the rapid growth and complexity of digital data. To this end, Large Language Models (LLMs) have emerged as powerful tools capable of automating and enhancing various digital forensic tasks, significantly transforming the field. Despite the strides made, general practitioners and forensic experts often lack a comprehensive understanding of the capabilities, principles, and limitations of LLM, which limits the full potential of LLM in forensic applications. To fill this gap, this paper aims to provide an accessible and systematic overview of how LLM has revolutionized the digital forensics approach. Specifically, it takes a look at the basic concepts of digital forensics, as well as the evolution of LLM, and emphasizes the superior capabilities of LLM. To connect theory and practice, relevant examples and real-world scenarios are discussed. We also critically analyze the current limitations of applying LLMs to digital forensics, including issues related to illusion, interpretability, bias, and ethical considerations. In addition, this paper outlines the prospects for future research, highlighting the need for effective use of LLMs for transparency, accountability, and robust standardization in the forensic process.}

\keywords{Large Language Model, Digital Forensics, Artificial Intelligence, Forensic Investigations}



\maketitle

\section{Introduction}

Digital forensics is a critical component in modern investigative and judicial processes, which involve the systematic collection, analysis, and preservation of digital evidence from electronic devices and online activities~\cite{wickramasekara2025exploring,rahman2024leveraging,xu2024transforming}. Its primary objective is to uncover factual information related to cybercrimes, fraud, unauthorized access, and other illicit activities~\cite{rogers2006two}. Digital forensics has played a pivotal role in solving high-profile cybercrime cases. For example, in the 2014 Sony Pictures hack, forensic investigators traced the breach back to North Korean hackers, who leaked confidential company data, emails, and unreleased films as part of a geopolitical cyber attack~\cite{ismail2017sony}. The investigation relied on digital forensics techniques such as analyzing network logs, identifying malware signatures, and attributing IP addresses to suspected attackers. As another example, in the 2016 Democratic National Committee (DNC) email leak, digital forensic experts identified sophisticated spear-phishing tactics and linked the attack to Russian-backed hacking groups, influencing the U.S. presidential election~\cite{marmura2018wikileaks,confessore2016hacked}. Beyond cyber espionage, digital forensics has also been crucial in financial fraud investigations. For instance, the Silk Road darknet marketplace, a notorious online black market, was dismantled in 2013 through extensive forensic analysis of Bitcoin transactions, server logs, and encrypted messages~\cite{minnaar2017online,lacson201621st}. Forensic experts traced Bitcoin payments to the marketplace's operator, Ross Ulbricht, ultimately leading to his arrest and life sentence. In another case, the Enron scandal saw digital forensics specialists recover crucial deleted emails and financial records, providing key evidence in one of the largest corporate fraud investigations in history~\cite{negangard2020electronic}. Additionally, digital forensic methodologies have been instrumental in child exploitation cases, where law enforcement agencies track online predators by analyzing metadata in images, chat logs, and digital footprints left on the dark web~\cite{kim5110258digital,quick2018digital}.

These case studies highlight the effectiveness of digital forensics in various domains, but they also demonstrate how investigators increasingly encounter complex technological challenges that test the limits of current methodologies. The primary issue is that traditional digital forensic techniques predominantly rely on manual or semi-automated approaches, requiring intensive human involvement~\cite{chen2014cloud}. These methods suffer from several inherent limitations. Firstly, they are labor intensive and time consuming, making them less effective in handling large-scale and sophisticated cyber incidents~\cite{fernando2023multidimensional}. Secondly, traditional methods often struggle to maintain consistency and accuracy due to human error and subjective judgments, potentially compromising evidence reliability. In addition, conventional forensic tools exhibit limited adaptability to evolving cyber threats, and their capability to identify complex interrelationships among evidence entities remains constrained~\cite{malik2024cloud,garach2024comprehensive}. 

 \begin{figure}[h]
    \centering
    \includegraphics[width=1\textwidth]{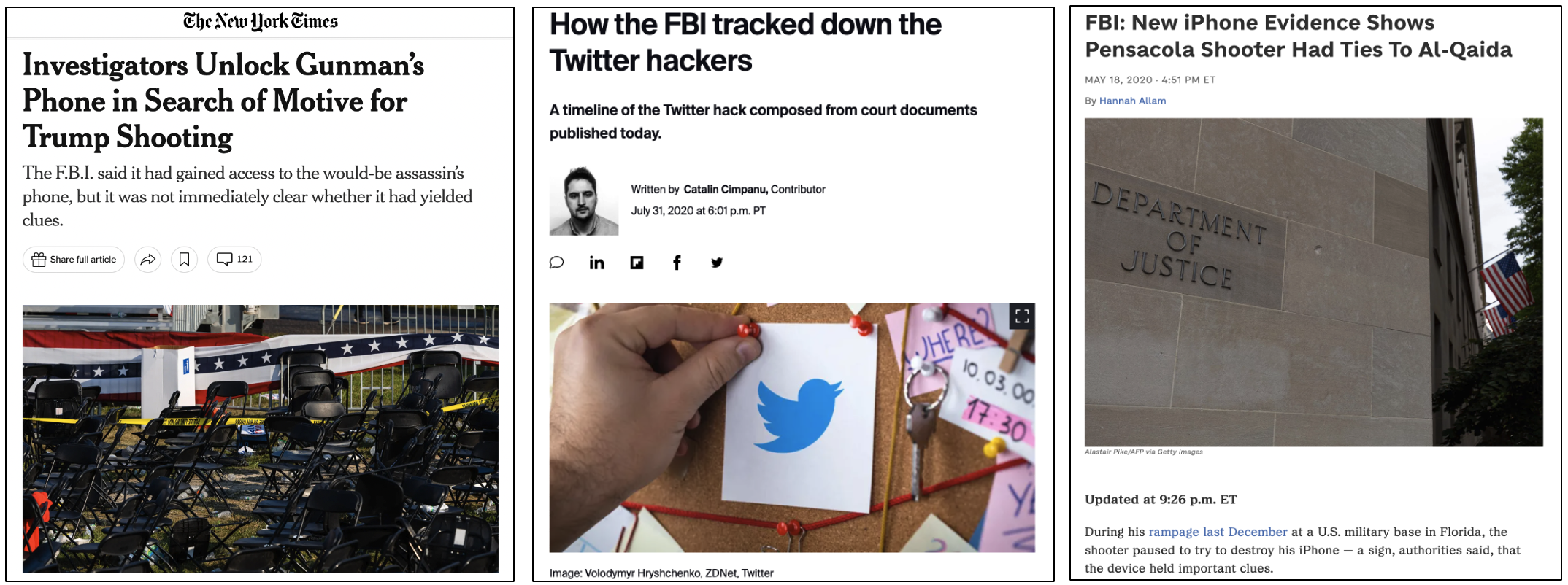}
    \captionsetup{justification=centering}
    \vspace{0.1cm}
    \caption{Investigation of real-world digital forensics cases in recent years.}
    \label{fig:realworld}
\end{figure}

\nocite{wang2023preventing, zhang2023individual, wang2023fg2an, yazdani2024comprehensive, wang2023mitigating, chinta2023optimization, wang2024history, chu2024fairness, dzuong2024uncertain, yin2024improving, wang2023towards, wang2024toward, chinta2024fairaied, doan2024fairness, chinta2024ai, wang2024individual1, doan2024fairness1, wang2024advancing}

Several recent cases, as shown in Figure~\ref{fig:realworld} illustrate these limitations. For instance, in July 2024, a gunman attempted to assassinate former U.S. President Donald Trump during a public rally, prompting an intensive investigation by federal authorities~\cite{swansonbullets,nyt_trump_shooting_2024}. Following the suspect’s capture, the FBI conducted a forensic analysis of his mobile device to uncover potential motives, affiliations, and premeditated plans. Investigators faced significant challenges in bypassing the phone’s security mechanisms, including encryption and biometric locks. Once access was gained, digital forensic teams meticulously analyzed call logs, text messages, encrypted messaging apps, and location history. Additionally, they examined the suspect’s social media interactions, online searches, and affiliations with extremist groups. Traditional analysis methods required extensive manual effort to filter through vast amounts of digital data, cross-reference communication patterns, and verify links between different sources. Such challenges were similarly highlighted in the 2020 Twitter cryptocurrency scam, where cybercriminals compromised multiple high-profile accounts to solicit Bitcoin payments~\cite{bartoletti2021cryptocurrency}. The FBI’s digital forensic teams encountered substantial hurdles as they manually cross-referenced Discord chat logs, leaked hacker forum databases, cryptocurrency wallet transactions, and IP addresses to identify the perpetrators. Although successful, this method revealed critical shortcomings in efficiently correlating and interpreting multi-dimensional digital evidence streams, demonstrating the urgent need for more advanced forensic capabilities~\cite{cimpanu2020fbi}. These complexities also surfaced prominently in the FBI’s investigation into the Pensacola naval base shooting in 2019~\cite{kessler2020cryptography,clarke2020pensacola}. In that case, the assailant’s encrypted iPhones had sustained physical damage, and Apple refused official requests for access assistance. Consequently, FBI forensic experts spent months painstakingly repairing hardware and circumventing encryption to retrieve data. Eventually, the recovered digital evidence established clear connections between the attacker and foreign terrorist entities. However, the prolonged investigative timeline underscored limitations inherent in traditional forensic methodologies when handling encrypted devices and fragmented digital traces~\cite{vaghela2024digital,allam2020pensacola}. Collectively, these cases emphasize the increasing necessity of integrating AI-driven digital forensic tools. Leveraging automation, intelligent data analysis, and advanced pattern recognition technologies could significantly enhance investigative speed, consistency, and accuracy, effectively addressing the growing scale, complexity, and sophistication of contemporary cyber threats~\cite{nayak2024ai,akeiber2025comprehensive}.

To address these substantial challenges in digital forensics, recent advances in artificial intelligence offer promising solutions. Notably, large language models (LLMs), such as the generative pre-trained transformer (GPT) series of models and the Gemini series, have emerged as powerful tools with the potential to transform digital forensic practices~\cite{liu2024toward,wang2024history,doan2024fairness,chu2024fairness}. These advanced AI models are designed to understand, interpret, generate and analyze human language with unprecedented accuracy. Using vast amounts of textual data from diverse sources, LLMs exhibit exceptional capabilities in natural language processing, pattern recognition, and semantic understanding~\cite{jin2023rethinking,ferrag2025generative,yao2024survey,valmeekam2022large}. Their ability to extract meaningful insights from large, unstructured datasets makes them invaluable in digital forensic investigations~\cite{kumarage2024survey}.

The application of these sophisticated models directly addresses the limitations of traditional forensic approaches identified earlier. One of the most impactful applications of LLMs in forensic analysis is their ability to automate and streamline the evidence identification process.~\cite{ahmed2016survey,liu2024large} Traditional methods require investigators to manually sift through enormous volumes of text, such as emails, chat logs, social media posts, and financial records. LLMs, on the other hand, can swiftly process and categorize these texts, recognizing patterns, detecting anomalies, and identifying crucial connections between disparate pieces of evidence~\cite{velasco2022cybercrime,mijwil2023towards,zhang2023forensiq}. This capability significantly accelerates investigative timelines while reducing the risk of human error~\cite{siddiqui2018application}. In the aforementioned high-profile investigation, integrating LLM-powered analytical tools could have played a transformative role in expediting the investigative process. By rapidly categorizing and interpreting textual evidence, LLMs can highlight potential leads, uncover hidden relationships, and help investigators piece together a cohesive narrative~\cite{chen2024exploring,yang2024harnessing,smirnov2025enhancing}. Moreover, their ability to process multilingual content ensures that forensic teams can analyze communication in different languages and cultural contexts without the need for extensive translation efforts~\cite{kao2025accelerating,karie2019diverging}. LLMs also improve forensic data interpretation by facilitating the reconstruction of complex evidence relationships. They can map connections between personal identifiers, such as names, addresses, and phone numbers, and correlate them with network activity, financial transactions, and geolocation data~\cite{arshad2018digital,klasen2024invisible,daniel2011digital}. This holistic approach allows investigators to establish links between suspects, victims, and illicit activities with greater precision~\cite{caballero2024leveraging}. Another crucial advantage of LLMs in digital forensics is their ability to handle large-scale data integration. Digital evidence is often scattered across multiple sources, including cloud storage, encrypted messaging platforms, and darknet forums. LLMs, combined with knowledge graph techniques, can aggregate and visualize these fragmented data points, making it easier to identify trends, associations, and key actors within an investigation~\cite{akhtar2025llm}.

While the benefits of LLMs for digital forensics are substantial, their implementation is not without challenges that need to be carefully considered~\cite{labajova2023state,khlaif2023potential,raza2024ai}. In addition, despite their transformative potential, the adoption of LLMs in forensic investigations also introduces new challenges, including concerns over interpretability, bias, and the reliability of AI-generated insights~\cite{azodi2020opening,quang2025insight}. Ensuring transparency in forensic AI applications is crucial to maintaining credibility in judicial proceedings~\cite{wischmeyer2019artificial}. Therefore, addressing these concerns through clear guidelines, rigorous validation procedures, and transparent reporting practices becomes essential~\cite{djeffal2019artificial,cath2018governing,baror2021natural}. To this end, this paper explores how LLMs can fundamentally change digital forensics practices by automating evidence analysis, extracting insightful information, and enhancing the judicial process, and attempts to provide a comprehensive understanding of the practical applications, potential limitations, and broader implications of LLMs in digital forensics investigations~\cite{jain2024enhancing}.

\noindent \textbf{Paper Structure.}
The subsequent sections of this paper are structured as follows: Section 2 introduces foundational concepts and highlights the limitations of training-based ai digital forensic methodologies. Section 3 details the principles and capabilities of large-scale language modeling and presents practical applications and real-world case studies. Section 4 evaluates the current challenges and limitations faced when deploying LLMs in forensic scenarios. Finally, Section 5 discusses opportunities and directions for future research.

\section{Fundamentals of Digital Forensics}
% 1. Definition , Goals

% 2. Key Evidence Types

% 3. Evidence Relationships
This section examines in depth the core principles of digital forensics. Understanding these fundamentals is critical to grasping how modern investigative processes utilize digital evidence to combat cybercrime, fraud, and other illicit activities.

\nocite{wang2024group, wang2024individual, yin2024accessible, wang2025fg, wang2025graph, wang2025fair, wang2025towards, zhang2024fairness, zhang2024inpractice, zhang2019faht, zhang2021fair, zhang2022longitudinal, zhang2023censored, zhang2025fairness, zhang2023fairness,zhang2021disentangled}

\subsection{Definition and Goals}
Digital forensics is the systematic process of identifying, collecting, preserving, analyzing, and presenting digital evidence in a legally admissible manner~\cite{casey2011digital,walker2001digital}. It is widely used in criminal investigations, cybersecurity incidents, corporate fraud detection, and other legal proceedings\cite{sharevski2015rules}. The primary goal of digital forensics is to uncover and reconstruct events related to cybercrimes, unauthorized access, financial fraud, intellectual property theft, and other illicit digital activities. By leveraging digital forensic techniques, investigators can retrieve hidden, deleted, or encrypted data to support legal actions and improve cybersecurity measures~\cite{ademu2011new}.

\subsection{Digital Forensic Evidence Entities}

A digital forensic evidence entity represents the smallest indivisible unit of digital information possessing forensic significance. Such entities serve as fundamental building blocks for reconstructing digital events and verifying their authenticity. These entities are categorized according to their functional purpose in supporting investigative analysis~\cite{quick2014data}. Table~\ref{tab:entites} provides a detailed description of these entities by functional purpose. The Content-Descriptive Entities help investigators understand the nature, source, or intended use of digital artifacts, providing essential context to evidence collected during an investigation~\cite{quick2014impacts}. In contrast, auxiliary entities supplement this understanding by offering validation, verification, and support to primary descriptive evidence, ensuring the reliability and integrity of forensic findings.

\begin{table*}[h]
\centering
\caption{Functional Categories and Descriptions of Digital Forensic Evidence Entities}
\label{tab:entites}
\begin{adjustbox}{scale=1}
\begin{tabular}{>{\raggedright\arraybackslash}p{0.8in}|>{\centering\arraybackslash}m{1.4in}|>{\raggedright\arraybackslash}m{3.2in}}
\hlineB{2}
\textbf{Categories} & \textbf{Digital Forensic Evidence Entities} & \textbf{Description}  \\
\cline{1-3}
\multirow{2}{*}{ \parbox{0.8in}{Content-Descriptive}} & File Names & Identifiers given to files, potentially revealing their content, origin, or intended purpose.\\
\cline{2-3} 
\multirow{1}{*}{ Entities} &IP Addresses & Numerical labels assigned to devices on a network, crucial for tracking and attributing online activities.  \\
\cline{1-3} 
\multirow{2}{*}{Auxiliary } &Timestamps & Specific points in time indicating events such as file creation, modification, or access. \\
\cline{2-3} 
\multirow{1}{*}{ Entities} &Hashes & Unique identifiers generated from data content used to verify file integrity and detect tampering. \\
\cline{1-3}

\hlineB{2}
\end{tabular}
\end{adjustbox}
\end{table*}

\subsection{Key Evidence Types}
While the theory of digital forensic evidence entities is understood based on their functional role, real-world forensic investigations often require more specific categorization.  Investigators routinely encounter various forms of digital evidence, which must be clearly identified and categorized to effectively address complex forensic challenges~\cite{lillis2016current,vincze2016challenges}. The following summarizes specific categories of digital evidence, which are classified according to their relevance and investigative role, and demonstrates how the theoretical framework translates into operational forensic practice~\cite{rowlingson2004ten}.

\begin{itemize}
\item \textbf{Personal Identifiers:} Names, addresses, phone numbers, email addresses, social security numbers, and other personal information. In identity theft cases, stolen personal identifiers are typically located within phishing emails, compromised databases, or fraudulent registrations.

\item \textbf{Network Information:} IP addresses, MAC addresses, login credentials, and network logs crucial for tracing user actions across devices and networks. Investigators often utilize this data to pinpoint sources of unauthorized access, as exemplified by numerous cases of insider threats and external intrusions.

\item \textbf{Communication Records:} Emails, text messages, social media messages, and call logs that capture interactions among individuals or groups. Analysis of these records has been pivotal in solving cases involving cyberbullying, insider trading, and organized crime.

\item \textbf{Financial Data:} Bank account details, credit card transactions, cryptocurrency wallet addresses, and transaction histories essential in tracking financial fraud and money laundering. Forensic analysts frequently exploit blockchain technology to unravel cryptocurrency-based criminal networks.

\item \textbf{Location Data:} GPS coordinates, timestamps, and geolocation logs, enabling investigators to track movements and verify alibis. This form of data has notably been employed in criminal cases where mobile device locations provided critical evidence linking suspects to crime scenes.

\item \textbf{Internet Activity:} Web browsing history, search queries, downloads, and online interactions offering deep insights into user behaviors and intentions. These digital footprints have been invaluable in cases involving radicalization, online harassment, and cyberstalking.

\item \textbf{File Metadata:} Information including timestamps, file paths, and document version histories, useful for establishing file authenticity and tracking document manipulation. Metadata analysis has been critical in corporate espionage investigations and cases of intellectual property theft.

\item \textbf{Device Logs and System Artifacts:} System event logs, registry entries, and application usage records, offering detailed insight into user activities and system states. In investigations of data breaches or corporate sabotage, these logs have provided evidence disproving fabricated user accounts and narratives.
\end{itemize}

\subsection{Evidence Relationships}
Having discussed specific categories of digital evidence encountered in practical forensic scenarios, it is crucial to recognize that these pieces of evidence rarely exist in isolation. Instead, they form intricate networks of relationships that significantly enhance investigative analysis~\cite{amato2017correlation}. Understanding these interconnections allows investigators to reconstruct detailed narratives, establish causality, and verify the authenticity of digital evidence comprehensively~\cite{horsman2024importance}. The key relationship categories include:

\textbf{i) Contextual Relationships:} These relationships provide situational context, helping investigators understand the origin, purpose, or usage of evidence. For example, linking file names to their content or correlating an IP address to a geographical location helps determine the source and intention behind cyber incidents.

\textbf{ii) Causal Relationships:} Highlight cause-and-effect dynamics between evidence entities. Identifying the correlation between an IP address and a specific time stamp can establish a suspect's direct involvement in unauthorized access or data manipulation.

\textbf{iii) Associative Relationships:} Connect seemingly independent evidence through shared attributes. Similar file hashes detected across multiple devices may suggest deliberate data duplication, exfiltration, or manipulation efforts by malicious actors.

\textbf{iv) Communication Relationships:} Reveal interaction patterns among individuals or systems. Analyzing communication logs such as phone records, emails, or chat messages has proven essential in dismantling criminal networks, uncovering collaboration among perpetrators, and mapping complex interactions in cybercrime investigations.

\textbf{v) Ownership and Association:} Establish explicit connections between individuals and digital devices, accounts, or data. Digital forensic efforts routinely involve associating specific devices or accounts with suspects, thereby strengthening investigative narratives and courtroom presentations.

\textbf{vi) Temporal Relationships:} Establish a chronological sequence or simultaneity of events. Timestamp analysis enables forensic examiners to confirm or refute suspect claims, authenticate alibis, and determine exact timelines of incidents, especially critical in high-stakes criminal and corporate investigations.

\subsection{Limitation Of Training-based AI For Digital Forensics}
While AI driven methodologies offer significant advancements in digital forensic investigations, several inherent limitations constrain their effectiveness, particularly when employing training-based AI approaches:

\textbf{i) Data Scarcity:}
Obtaining sufficient and diverse training data representative of real-world cyber incidents poses significant challenges. Often, the available data is limited to specific case types, such as addresses extracted predominantly from certain criminal activities like shootings. This lack of comprehensive and varied datasets can severely restrict the AI model's ability to generalize across different forensic scenarios~\cite{ahmed2016survey}.
 
\textbf{ii) Data Pre-processing Challenges:} Even seemingly simple tasks, such as identifying addresses using Named-Entity Recognition (NER), introduce considerable pre-processing complexity before AI models can be effectively applied. These tasks often require multiple pre-processing steps, including expanding abbreviations (e.g., converting``St.'' to``Street''), standardizing formats (e.g., ``123 Main St Apt 4B'' to ``123 Main Street, Apartment 4B''), normalizing state names (e.g., ``California'' to ``CA''), and removing extra whitespace (e.g., converting ``456 Elm St'' to``456 Elm St.''). These additional pre-processing steps significantly increase the complexity, time, and resources required, underscoring the limitations associated with direct training-based AI approaches in digital forensic analyses.

\textbf{iii) AI Models Lack Adaptability:}
AI models developed for digital forensic tasks are typically designed and optimized for specific, narrowly defined functions. For instance, an AI model trained explicitly for recognizing addresses will likely exhibit limited performance when tasked with identifying other types of information, such as personal names or financial records~\cite{ferrag2024generative}. This specialized training makes it challenging to apply these models broadly across the diverse range of forensic tasks investigators encounter.

\textbf{iv) Difficulty in Extracting Evidence Relationships:}
Identifying and analyzing the numerous intricate relationships among digital evidence entities is inherently complex. Training-based AI methods often struggle to capture the full depth and nuance of these interactions, given the extensive variety and subtlety in relationships, including contextual, causal, associative, communication-based, ownership-based, and temporal connections~\cite{kucharavy2023fundamentals}. Consequently, traditional training-based approaches may not recognize critical evidence correlations, potentially undermining the accuracy and comprehensiveness of forensic analyzes.

\section{Large Language Models for Digital Forensics}

\subsection{Why LLMs For Digital Forensics}

Large Language Models (LLMs) represent a sophisticated category of artificial intelligence models, primarily designed to understand, generate, and interact with natural language text~\cite{zhao2023survey}. These models typically utilize deep learning architectures, such as transformers, which rely on self-attention mechanisms to capture intricate contextual relationships within textual data~\cite{chang2024survey}. The development and training of LLMs involve vast datasets, often comprising billions of words, enabling these models to acquire a deep understanding of syntax, semantics, and contextual nuances inherent in human languages.

One of the most prominent examples of LLMs is the Generative pre-trained Transformer (GPT) series developed by OpenAI, including GPT-3 and GPT-4. These models exhibit exceptional capabilities across a wide range of natural language processing (NLP) tasks, such as text generation, summarization, translation, sentiment analysis, question answering, and entity extraction~\cite{naveed2023comprehensive,shanahan2024talking}. Their impressive versatility stems from their ability to capture long-range contextual dependencies and their extensive training on diverse textual resources such as websites, books, articles, and other publicly available information.

The LLMs training process generally involves two main phases: pre-training and fine-tuning. During pre-training, models are exposed to vast, unsupervised text corpora, learning general language patterns, syntax, and semantic relationships without specific task-oriented labels. In the fine-tuning phase, LLMs are further trained in task-specific datasets, adapting their general language comprehension skills to effectively perform targeted NLP tasks~\cite{zhao2024explainability}. This two-phase approach significantly enhances their adaptability and performance across diverse domains.

LLMs are trained on vast amounts of text data and exhibit exceptional capabilities in learning linguistic patterns, structural semantics, and contextual dependencies. These attributes make them uniquely suited for applications in digital forensics, where the volume and heterogeneity of digital evidence can overwhelm traditional analysis techniques. In digital forensic investigations, evidence often exists in unstructured or semi-structured forms, such as chat logs, emails, file metadata, browsing histories, and system logs. Manually extracting meaningful patterns or relationships from such data is labor intensive and time consuming~\cite{chen2021evaluating}. LLMs can assist by automatically identifying named entities, classifying document types, summarizing lengthy communication threads, detecting suspicious patterns, and establishing semantic links across diverse artifacts.

LLMs offer the ability to generalize from limited context, which is particularly useful in forensic settings where fragmented or incomplete evidence is common. Their pre-training on diverse data sources also allows them to recognize and interpret technical jargon, code snippets, and colloquial expressions, enabling them to analyze evidence drawn from varied digital environments~\cite{zhao2023survey}. LLMs also support multi-turn interactions, allowing investigators to iteratively refine queries or extract context-sensitive information from large datasets in a conversational manner. This interaction paradigm not only enhances usability but also reduces the need for technical expertise in formulating complex forensic queries.

Therefore, LLMs have great potential to enable digital forensics given their training on large textual datasets and powerful pattern learning capabilities, and by utilizing LLms' advanced linguistic understanding analytical and generative capabilities, key information can be extracted from large amounts of data, significantly streamlining the analysis of evidence, enhancing the process of informed decision-making, and improving overall forensic outcomes~\cite{chang2024survey}.

\subsection{LLMs-driven methods in Digital Forensics}
Integrating LLM into forensic workflows has emerged as a promising approach as investigators seek new tools to improve the accuracy, efficiency, and scalability of their analyses. This section provides an overview of current methodological frameworks and empirical case studies in which LLM has been effectively utilized in digital forensic environments. Specifically, it highlights representative applications, evaluates their practical effectiveness, and identifies methodological insights that have emerged from these real-world implementations.

\subsubsection{LLM-driven Construction of Evidence Networks}
Utilizing their powerful pattern recognition and relationship extraction capabilities, LLMs introduce innovative methods to improve the efficiency and accuracy of forensic investigations~\cite{zhoullm}. One prominent example of such LLM-driven methods involves utilizing GPT-4-turbo to systematically identify and visualize patterns within digital forensic evidence. This method constructs a structured graph G = (V, E), where nodes (V) represent individual evidence items—such as names, addresses, and phone numbers—while edges (E) depict relationships connecting these items. Each edge is labeled explicitly to describe the nature of the connection, such as ``owns'' for ownership (a person owning a phone number) or ``lives-in'' for residency (a person residing at an address). These clearly defined relationships enable the creation of comprehensive visual representations that simplify the analysis of intricate forensic data.

\begin{figure}[h]
    \centering
    \includegraphics[width=1\textwidth]{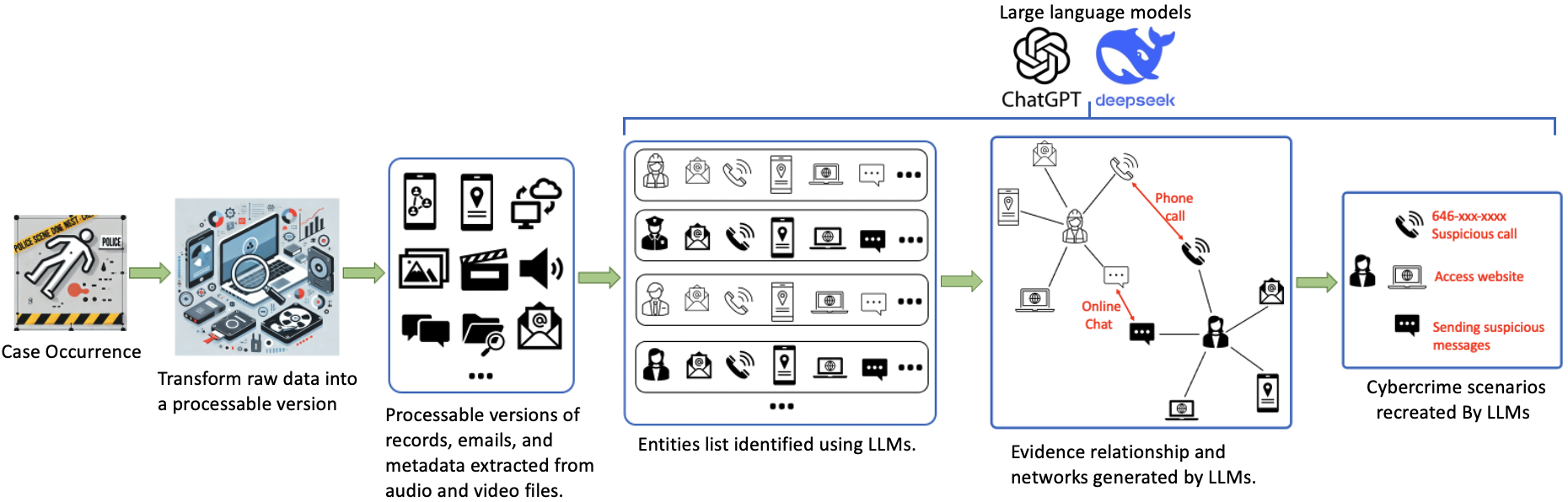}
    \vspace{0.1cm}
    \caption{A LLMs Methods to Understanding Cybercrime via Evidence Networks.}
    \label{fig:llms-driven}
\end{figure}

The main procedural steps shown in Figure~\ref{fig:llms-driven} typically include:

\textbf{i) Transform raw data into a processable versions: } This step involves extracting and standardizing evidence from mobile devices, personal electronic devices and especially from their embedded multimedia card storage. Considering that these devices often contain fragmented, hidden or deleted data in binary form, converting this information into a clear text format is essential for the accurate analysis of llm.

\textbf{ii) Identifying evidence entities and their relationships:} Researchers create and test tailored prompts that guide LLMs in systematically extracting relevant evidence entities from structured textual data such as chat logs and system records. A representative prompt could be: ``Act as an experienced digital forensic investigator. Extract evidence entities like names, addresses, and phone numbers from the given text and outline the relationships among these entities.''

\textbf{iii) Constructing evidence networks:} This step involves connecting isolated pieces of evidence to form coherent networks. Connections are identified based on proximity, either physical (line distance in text) or semantic (inferred through LLMs), under the assumption that closely positioned entities are likely interrelated.

\textbf{iv) Deriving insights into criminal behavior:} Lastly, these constructed evidence networks are analyzed to uncover significant insights into criminal activities, behaviors, and underlying relationship patterns. This detailed examination of interconnected evidence provides forensic investigators with critical information that enhances their understanding of complex criminal scenarios.

\subsubsection{LLM-driven Invocation Log Analysis for Digital Forensics}
Chernyshev \textit{et al}. proposed a novel forensic methodology aimed at detecting prompt injection attacks in applications integrated with LLMs~\cite{chernyshev2023towards}. The core innovation of this approach lies in leveraging invocation logs,a structured records of LLM interactions, as a primary evidentiary source for digital forensic investigations.

Their method involves constructing a simplified yet representative experimental scenario that emulates real-world LLM-integrated web applications, and Figure~\ref{fig:log} illustrates its workflow. Specifically, the authors developed a web-based application utilizing GPT-3.5 via the LangChain framework. In this scenario, users' natural language queries are converted by the LLM into Structured Query Language (SQL) statements, subsequently executed against a backend relational database. To create realistic attack conditions, the authors manually designed a set of malicious prompts to simulate direct prompt injection attacks, such as dropping database tables or bypassing access control restrictions.

To facilitate digital forensic readiness (DFR), the authors introduced structured logging mechanisms, termed LLM invocation logging, into their experimental system. Each invocation log entry captured essential forensic metadata including a timestamp, unique request identifier, input prompt (user’s query), and corresponding LLM output, generating structured JSON-formatted logs, thereby ensuring traceability and forensic integrity.

\begin{figure}[h]
    \centering
    \includegraphics[width=1\textwidth]{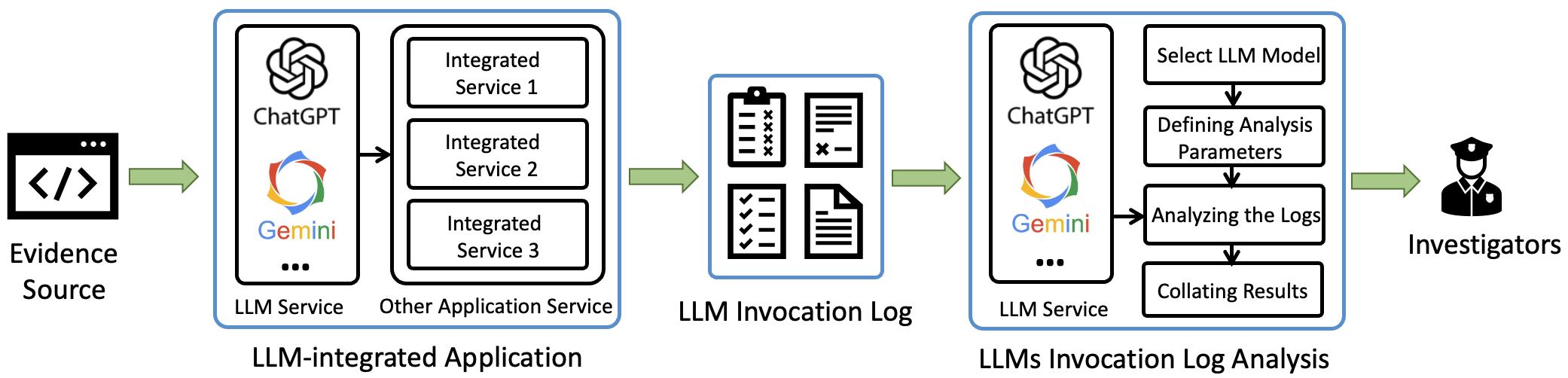}
    \vspace{0.1cm}
    \caption{Digital Forensic Analysis Workflow with LLM Invocation Logs.}
    \label{fig:log}
\end{figure}

For forensic analysis, the collected invocation logs were processed using an active analysis strategy in which multiple contemporary LLMs acted as forensic analysts. Given that different models have significantly varying context windows—for instance, from 8,182 tokens for llama3-70b-instruct to 1 million tokens for gemini-1.5-flash and gemini-1.5-pro, the authors evaluated analysis approaches both with models capable of accepting the entirety of the invocation logs as input and those requiring splitting log entries into smaller window chunks for sequential processing. Their analysis involved four main steps: \textbf{i) Selecting an LLM model for analysis,} GPT-4, Gemini or other similar models; \textbf{ii) Defining key analysis parameters,} such as the LLM's temperature and context window size; \textbf{iii) Actively analyzing the logs with the chosen configuration,} using the chosen model and parameters; \textbf{iv) Collating the results,} summarizing key findings and observations. These steps were systematically repeated until all desired combinations of models and analysis parameters had been evaluated. Unlike previous works exploring LLM usage for anomaly detection that employed pre-summarization, this approach solely relied on active log analysis without context summary creation. 
Specifically, each model was provided log entries within a predefined context window, accompanied by instructions to identify potential security incidents and articulate justifications. The models returned structured JSON outputs indicating detection decisions (either ``NORMAL” or ``INCIDENT”), suspicious log indices, and descriptive reasoning. This direct approach significantly reduced the overall number of calls to the LLM, consequently decreasing both the total time required for log analysis and the potential cost.

This approach illustrates important advances in digital forensic readiness for LLMs-drivenn systems, showing how invocation log analysis performed by the LLM itself can provide practical forensic capabilities for identifying sophisticated hint injection attacks.

\subsubsection{LLM-driven Mobile Evidence Contextual Analysis }
Kim \textit{et al}. propose a comprehensive and operationally grounded framework for mobile forensics, termed Mobile Evidence Contextual Analysis(MECA)~\cite{kim5110258digital}. This framework addresses the practical challenges law enforcement faces in analyzing large volumes of mobile messenger data, particularly under tight legal time constraints. Rather than relying solely on traditional keyword-based filtering, MECA leverages the contextual reasoning capabilities of LLMs to infer the presence of criminal intent or activity embedded in ambiguous or euphemistic language. The method is notable not only for its application to real-world forensic data but also for its holistic integration of forensic tools, data pre-processing, and prompt engineering.

The framework begins with the acquisition of mobile communication data using professional forensic software tools. Specifically, the authors employ MD-NEXT for physical data extraction and MD-RED for data parsing and visualization. These tools support the collection of structured communication records from seized smartphones, which are exported in formats like CSV or Excel for downstream processing. To ensure compliance with privacy and ethical standards, all personal identifiers within the dataset are anonymized using Named Entity Recognition (NER), with supplementary masking strategies applied to phone numbers and emails to minimize reidentification risk.

\begin{figure}[h]
    \centering
    \includegraphics[width=1\textwidth]{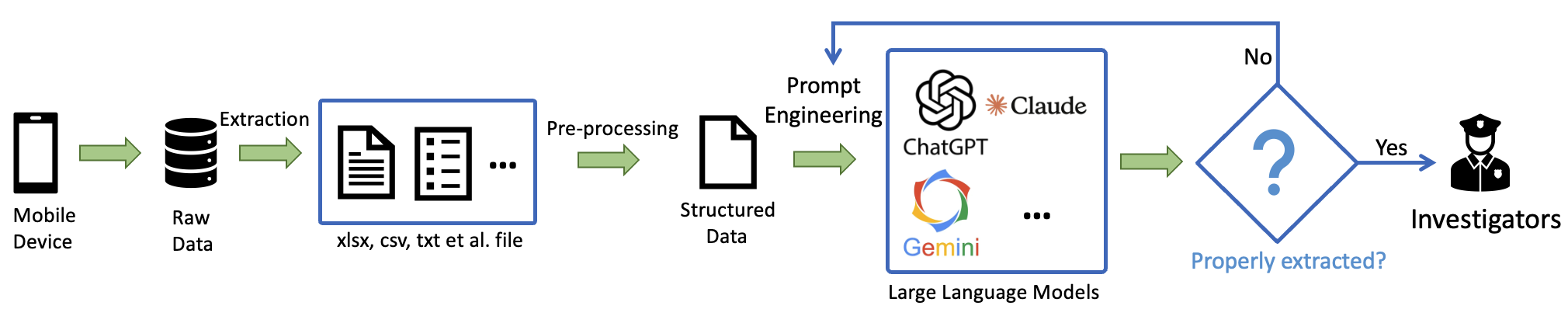}
    \caption{Overview of LLM-driven Mobile Evidence Contextual Analysis Framework.}
    \label{fig:log}
\end{figure}

Given the size and fragmented nature of mobile information logs, the authors introduced a pre-processing phase to construct coherent units of analysis appropriate for LLM input. This involves applying initial keyword filters, \textit{e.g.,} for terms such as``drugs”, to identify potentially relevant messages. In order to preserve conversational context, each filtered message is augmented with surrounding messages in the same chat window, typically 40 lines each before and after the targeted message. This produces a set of context-rich message fragments that reflect real-world communication patterns and facilitate semantic interpretation by the model.

Central to MECA’s effectiveness is its use of carefully crafted prompts to guide model behavior. Each prompt is designed to simulate the role of a forensic expert, instructing the model to evaluate whether a given message exchange is associated with criminal activity. The input is structured as key-value pairs, where the key represents the speaker and the value denotes the message content. Moreover, the authors implement a “Sandwich Prompting” technique—repeating instructions before and after the main content—to mitigate instruction forgetting, particularly in models like Gemini that may otherwise over-prioritize the input text.

Once the data and prompts are prepared, the framework employs three state-of-the-art LLMs, GPT-4o, Gemini 1.5, and Claude 3.5 to perform classification. Each model receives the structured conversational input and returns a binary judgment indicating whether the message set is relevant to the case. The authors also account for concerns around data privacy and model misuse by relying on commercial API deployments and explicitly documenting the privacy policies of each LLM provider. The use of multiple models not only allows performance benchmarking across architectures but also sets the stage for ensemble decision-making.

\subsubsection{Forensic Analysis of Artifacts from Microsoft’s Multi-Agent LLM Platform}

In this work by Walker \textit{et al.} proposes a comprehensive methodology for conducting forensic analysis of AutoGen, Microsoft’s multi-agent LLM framework\cite{walker2024forensic}. As AutoGen enables autonomous agent collaboration for task planning and execution, the forensic analysis of such systems introduces novel challenges, particularly in identifying, interpreting, and attributing the artifacts generated through agent interactions. The proposed methodology responds to this gap by establishing a structured, multi-layered approach to detecting the presence and behavior of AutoGen on a target system.

At the core of their approach is the idea of tracing the forensic footprint of LLM-driven agent interaction across three major layers of analysis: memory, disk, and network. Rather than focusing on any single modality of artifact, the methodology adopts a layered perspective to capture both persistent and volatile traces of AutoGen's activity on a host system. The authors hypothesize that, despite the encrypted nature of LLM-server communication and the ephemeral memory handling of modern OSes, a composite view of system-level behavior can reveal meaningful patterns associated with LLM agent activity.

The interaction model analyzed in the study involves two LLM-based agents: a UserProxyAgent, simulating a user that initiates tasks and evaluates responses; and an AssistantAgent, responsible for task execution. These agents interact through a feedback loop where task instructions and responses are exchanged programmatically. This model mirrors real-world use of AutoGen for distributed task planning and problem solving, and raises questions around forensic observability, \textit{i.e.}, what traces of such interactions persist on a compromised or analyzed system.

\begin{figure}[h]
    \centering
    \includegraphics[width=0.9\textwidth]{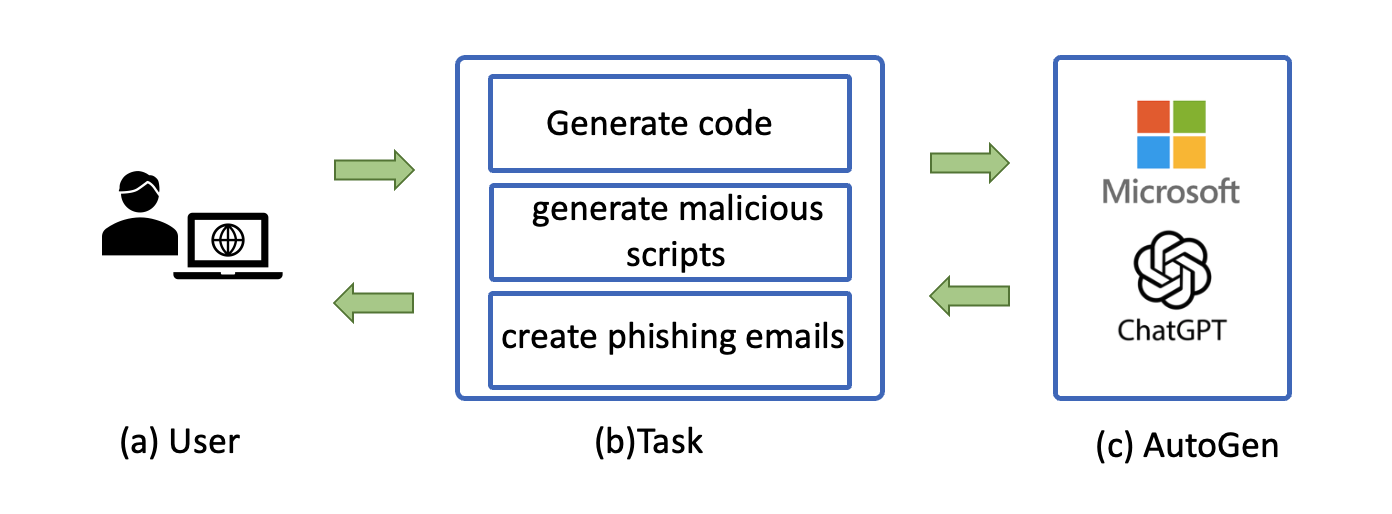}
    \caption{Overview of forensic analysis of AutoGen.}
    \label{fig:autogen-flow}
\end{figure}

To clarify this interaction workflow, Figure~\ref{fig:autogen-flow} illustrates the operational model used in the study. This process begins with a user operating from a local environment (A), where they initiate specific task prompts (B), such as generating code, crafting phishing emails, or producing malicious scripts. These prompts are passed to the AutoGen system (C), which coordinates interactions between LLM agents, typically a UserProxyAgent and an AssistantAgen, powered by models like GPT-3.5. The agents exchange messages programmatically until a task is completed, with AutoGen returning the model-generated output to the user. This controlled interaction loop is essential for generating forensic artifacts, the researchers are able to capture and later examine forensic artifacts across memory, disk, and network layers.

The forensic method involves isolating the key points where AutoGen interacts with the system or external services and mapping those to potential artifact locations. For instance, the UserProxyAgent's initial prompt, along with the AssistantAgent's responses, may be retained in memory buffers, cached in application files, or transiently recorded in system logs. The methodology accounts for the limited durability of such data and therefore incorporates the use of tools that can extract low-level system state information, \textit{e.g.,} RAM dumps, temporary configuration files, browser traces.

A notable component of the method is its treatment of agent attribution—attempting to distinguish whether a given artifact was created by a human, a machine, or a cooperative agentic process. This is a particularly novel challenge in LLM forensics, since traditional forensic signatures are often agnostic to the cognitive or computational origin of content. The methodology, therefore, considers semantic and behavioral cues, \textit{e.g.,} structure of prompt chains, repeated execution patterns, lack of GUI interaction, that may help differentiate machine-driven output from human-involved interaction.

Additionally, the approach integrates lightweight static analysis techniques, such as string extraction from memory and file systems, with dynamic signature correlation, such as identifying AutoGen-related modules in Python environments or connections to known LLM service endpoints. This hybrid approach helps mitigate the limitations of any single forensic strategy and provides a more comprehensive account of AutoGen’s presence and behavior on the system.

The method sets a foundation for future forensic analysis of autonomous LLM systems, especially as they become more modular, compositional, and capable of unsupervised behavior. It emphasizes the need for multi-perspective evidence gathering, cross-layer correlation, and a deeper understanding of agent-based software design in order to maintain accountability in increasingly AI-driven digital environments.

\subsubsection{The Local LLM-driven Framework for Digital Forensic }
While large language models (LLMs) have demonstrated remarkable capability across various natural language processing tasks, their application in sensitive domains such as digital forensics presents unique challenges, including concerns about data privacy, security, and the need for specialized domain knowledge. Moreover, reliance on cloud-based solutions can introduce vulnerabilities related to data confidentiality and compliance, prompting the need for locally deployable LLMs tailored specifically to forensic purposes. Addressing these critical issues, Sharma \textit{et al.} introduced ForensicLLM, a specialized, locally deployable large language model designed explicitly for digital forensic applications using a retrieval-augmented fine-tuning (RAFT) methodology~\cite{sharma2025forensicllm}.

\begin{figure}[!htb]
    \centering
    \includegraphics[width=0.99\textwidth]{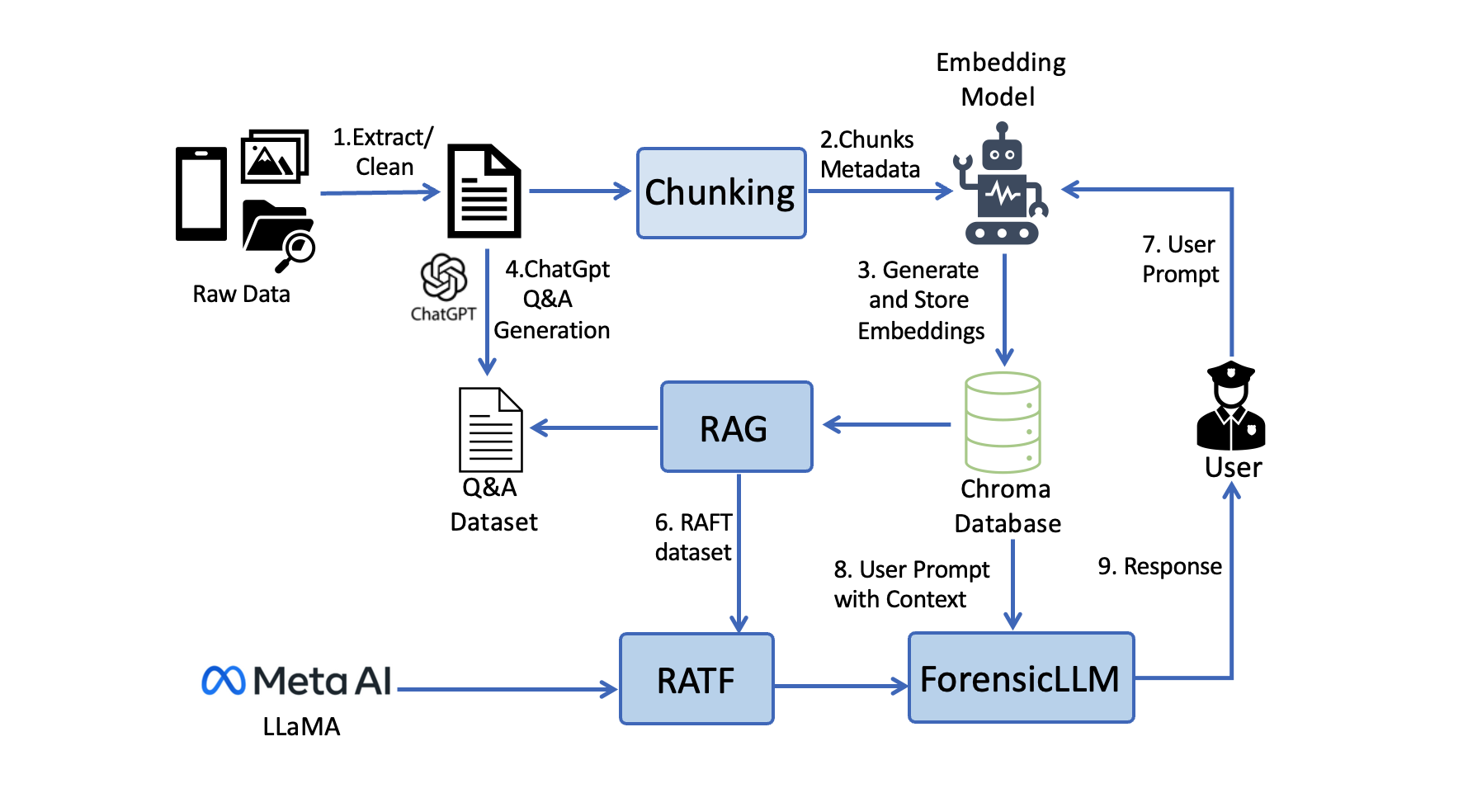}
    \caption{Overview of Retrieval-Augmented Fine-tuning (RAFT) for ForensicLLM.}
    \label{fig:local}
\end{figure}
Sharma \textit{et al}. utilized Meta's LLaMA-3.1–8B as the foundational model, enhancing it through fine-tuning with domain-specific content to address the unique reasoning demands inherent in digital forensic investigations. They began by compiling an extensive corpus comprising 1,082 peer-reviewed research articles sourced from the journal Forensic Science International: Digital Investigation, along with metadata extracted from 1,390 verified digital forensic artifacts obtained via the Artifact Genome Project. Textual contents from these research articles were segmented into semantically meaningful chunks of approximately 2,000 characters and embedded using the UAE-Large-V1 embedding model. Each chunk was enriched with associated metadata, including article titles and authors, with embeddings subsequently stored within a ChromaDB vector database to facilitate efficient retrieval during subsequent training and inference processes.

In the absence of suitable labeled question-answer datasets specific to digital forensic scenarios, the authors employed GPT-4 Turbo to generate approximately 10,000 synthetic question-answer pairs based directly upon the prepared literature corpus. This generation process was carefully guided using detailed prompting to ensure practically relevant, technically accurate content, maintaining faithful adherence to original source citations following APA standards.   

The fine-tuning procedure leveraged Quantized Low-Rank Adaptation, implementing a 4-bit quantization approach to optimize computational resource efficiency during training. Sharma \textit{et al.} adopted the Axolotl framework, utilizing standard practices such as cosine learning rate scheduling and early stopping based on validation set performance. During inference, ForensicLLM utilizes a retrieval-augmented generation (RAG) strategy, embedding user queries to dynamically retrieve relevant textual contexts from the vector database, which are then integrated into the model input to produce informed, verifiable, and accurate responses.

As shown in Figure~\ref{fig:local}, this figure outlines the sequence data processing and model-training pipeline, beginning with raw data extraction and cleaning, followed by segmentation into meaningful textual chunks. These text chunks, enriched with metadata, are then transformed into semantic embeddings using an embedding model and subsequently stored in a ChromaDB vector database. Simultaneously, synthetic Q\&A pairs are generated from the corpus using GPT-4 to form a structured training dataset. This Q\&A dataset is integrated with context retrieved from the vector database, forming the RAFT dataset utilized for fine-tuning the ForensicLLM model. Finally, during inference, user queries are embedded and matched with relevant contexts retrieved from Chroma datebase, enabling ForensicLLM to produce accurate, contextually informed, and traceable responses tailored specifically for digital forensic applications.

The retrieval-enhanced fine-tuning approach proposed by Sharma \textit{et al}. significantly impacts digital forensic practice by reducing common limitations associated with general-purpose language models, particularly hallucinations and factual inaccuracies. Their quantitative and qualitative evaluations demonstrated that ForensicLLM substantially improves response accuracy, relevance, and reliability, thus equipping forensic investigators with trustworthy, traceable analytical support capable of meeting rigorous evidentiary standards required in real-world forensic investigations.

% \subsection{Real-world Case study}

\section{Challenges and Limitations of Leveraging LLM in Digital Forensics}
The integration of large language models into digital forensics workflows has generated increasing interest due to their potential in automating documentation, evidence analysis, and decision support. However, their use also presents numerous challenges that arise both from the inherent properties of LLMs and from the specific requirements of forensic practice.

\subsection{LLM Inherent Challenges}
Several limitations are intrinsic to the architecture and training methodology of LLMs, which can hinder their safe and reliable deployment in forensic investigations.

%\noindent \textbf{i) Hallucinations}

\noindent \textbf{Hallucinations.} A prominent concern when employing LLMs is their propensity to produce hallucinated content—output that is grammatically coherent yet factually incorrect or fabricated. In the context of digital forensics, such inaccuracies can lead to the generation of false leads, thereby misguiding the investigation or introducing inadmissible evidence. For instance, in a controlled trial conducted by a cybersecurity firm, an LLM-generated case summary falsely inferred a link between an employee and a foreign contact based solely on contextual cues in a benign conversation log. This example highlights the necessity of human verification mechanisms prior to integrating LLM-generated information into forensic reports.

% \noindent \textbf{ii) Interpretability and Explainability}

\noindent \textbf{Interpretability and Explainability.} LLMs often exhibit poor explainability due to their black-box nature. While they can produce accurate results in many domains, their decision-making pathways are not transparent~\cite{zhao2024explainability}. This opacity becomes a critical issue in forensic analysis, where the rationale behind evidence interpretation must be traceable and defensible. In one documented instance during a civil litigation case, an LLM used in pre-trial discovery flagged certain emails as ``suspicious''; however, when the opposing counsel requested an explanation for these classifications, the legal team was unable to articulate the reasoning behind the model’s output. The lack of explainability ultimately led to the exclusion of the generated evidence.

% \noindent \textbf{iii) Lack of Domain-Specific Knowledge}

\noindent \textbf{Lack of Domain-Specific Knowledge.} General-purpose LLMs are trained on heterogeneous and largely non-specialized corpora. As such, they may not have the technical nuance necessary for forensic analysis. For example, when prompted to assess the contents of a memory dump, a widely used LLM erroneously flagged ``svchost.exe'' as malicious, failing to account for the legitimate role of the process in Windows systems. Such errors underscore the risk of applying unadapted LLMs in technical domains without appropriate domain fine-tuning.

% \noindent \textbf{iv) Bias and Fairness}

\noindent \textbf{Bias and Fairness.} Bias in LLMs-driven a reflection of the biases present in the training data—poses ethical and practical risks in forensic contexts~\cite{chu2024fairness}. Investigative results may be biased either by reinforcing existing stereotypes or by systematically prioritizing certain types of evidence. In a pilot study involving multilingual forensic datasets, an LLM-assisted classification system consistently deprioritized non-English chat logs, leading to a delay in the examination of relevant Arabic-language communications. This form of bias, if left unaddressed, could have far-reaching implications for fairness and due process in digital investigations~\cite{saxena2023missed}.

\subsection{Digital Forensics-Specific Challenges}

Although the inherent risks associated with LLMs pose general concerns in all domains, deploying these models within digital forensic workflows introduces additional challenges. Digital forensics imposes strict standards regarding evidence integrity, reproducibility, and procedural compliance, and these established forensic principles may conflict with the nature of LLM technologies. Consequently, integrating LLMs into digital forensic practices requires addressing specific challenges related to evidentiary standards, reproducibility, prompt sensitivity, standardization, and practitioner readiness.

% \noindent \textbf{i) Chain of Custody and Evidentiary Integrity}

\noindent \textbf{Chain of Custody and Evidentiary Integrity.} A core principle in forensic science is the preservation of chain of custody, that is, the ability to trace each step of evidence handling. When LLMs are employed, especially in cloud-based or third-party systems, questions arise regarding the preservation and auditability of evidence. In one European law enforcement case study, the use of an LLM to summarize mobile device contents inadvertently violated chain of custody procedures, as intermediate outputs were not systematically logged. As a result, the forensic findings were challenged on procedural grounds during judicial review.

% \noindent \textbf{ii) Non-determinism and Reproducibility}

\noindent \textbf{Non-determinism and Reproducibility.} Unlike deterministic forensic tools, LLMs are inherently probabilistic and may produce variable outputs even under identical input conditions. This variability undermines one of the key requirements of forensic science, namely reproducibility. In a university-led evaluation, an LLM used to reconstruct activity timelines from log data produced inconsistent event sequences across multiple runs. Such behavior poses serious threats to the reliability of forensic conclusions, particularly when outputs are used as part of expert witness testimony.

% \noindent \textbf{iii) Prompt Sensitivity}

\noindent \textbf{Prompt Sensitivity.} Related to non-determinism is the issue of prompt sensitivity, whereby subtle variations in phrasing can lead to significantly different model outputs. For instance, altering a prompt from “summarize suspicious behavior” to “summarize all activity” led an LLM to either omit or include key lateral movement indicators in the same dataset. The fragility of outputs based on minor linguistic changes necessitates rigorous prompt engineering and version control when using LLMs in evidentiary contexts.

% \noindent \textbf{iv) Lack of Standardization}

\noindent \textbf{Lack of Standardization.} There exists no established framework or industry-wide standard governing the use of LLMs in digital forensics. This absence of formal guidance has resulted in inconsistencies across investigative practices and raises concerns regarding admissibility and procedural fairness. In a simulated case involving two independent forensic teams, divergent conclusions were reached due to differences in prompt design, evidence filtering strategies, and LLM configurations. These discrepancies emphasize the need for standardized protocols and certification schemes for LLM-based forensic tools.

% \noindent \textbf{v) Training and Expertise Requirements}

\noindent \textbf{Training and Expertise Requirements.} The adoption of LLMs in digital forensic settings introduces new requirements for practitioner training. Investigators must possess not only technical forensic skills but also basic knowledge in AI, prompt design, and model validation. A field test conducted with junior investigators revealed that improper prompt use led to misclassifications of a legitimate mobile application as malicious, an error that could have been avoided with minimal training in AI reasoning mechanisms. The integration of LLMs thus demands a reevaluation of existing forensic training curricula to include AI literacy.

\section{Future Directions}
The intersection of large language models (LLMs) and digital forensics represents an emerging frontier with significant potential for transforming forensic investigations. Future research in this area promises to strengthen evidentiary integrity, promote greater accountability, and contribute broadly to societal trust and justice.

\subsection{Multi-Modal and Cross-Data Analysis}

Digital forensic investigations increasingly require holistic evidence interpretation across various data modalities—textual logs, network traffic, memory dumps, images, and audio. Emerging multi-modal LLMs (MLLMs) suggest promising capabilities for integrating diverse data forms into unified analytical frameworks. Integrating vision and language models could enable forensic assistants to analyze screenshots, correlate textual logs with visual artifacts, or interpret combined structured and unstructured forensic data. Research opportunities lie in developing robust multi-modal forensic LLMs capable of seamlessly analyzing multiple data types while maintaining accuracy across different modalities. Achieving this will require interdisciplinary collaboration and innovative design to bridge existing capability gaps.

\subsection{Explainability and Trust in LLM-Driven Analysis}

The inherently opaque reasoning of LLMs conflicts with forensic requirements for transparency and verifiability. Enhancing the explainability of LLM outputs is thus critical for building investigator trust. Future research should focus on methods that enable LLMs to justify their conclusions with explicit evidence references and step-by-step reasoning processes. Techniques like retrieval-augmented generation, where LLM outputs are grounded explicitly in input data and known forensic knowledge, can significantly improve credibility. Validation methods, such as cross-validation with multiple models or human-in-the-loop verification, should also be investigated to detect and mitigate errors and biases inherent in AI analyses.

\subsection{Domain-Specific LLMs Across Forensic Disciplines}

One crucial future direction involves the development of specialized, domain-specific LLMs tailored explicitly for various forensic applications such as memory forensics, malware analysis, network investigations, and log interpretation. General-purpose models typically lack the specialized technical understanding required to interpret detailed forensic artifacts accurately. Early examples, such as volGPT for memory analysis, have demonstrated the effectiveness of fine-tuned LLMs in accurately identifying ransomware processes while providing comprehensive explanations. Future research should systematically explore domain-specific models for forensic tasks, including artifact interpretation, filesystem analysis, and forensic triage. This specialization will necessitate creating dedicated forensic datasets, posing challenges related to data sensitivity and privacy that researchers must address through synthetic or anonymized datasets.

\subsection{Privacy and Legal Admissibility Challenges}

Integrating LLMs into forensic investigations raises significant privacy concerns and legal admissibility challenges. Public cloud-based solutions often conflict with chain-of-custody requirements, prompting the need for secure, offline LLM solutions deployable within forensic lab environments. Future research should focus on enhancing on-premise or federated AI models that preserve data confidentiality and comply with legal standards. Additionally, clearly defined legal frameworks and standards are needed for documenting and certifying AI processes, ensuring their outputs withstand judicial scrutiny. Collaborative research among technologists, legal scholars, and policymakers is necessary to bridge these gaps and ensure that LLM-assisted forensic analyses meet rigorous evidentiary standards.

\subsection{Integration with Traditional Forensic Tools and Workflows}

Future research must explore the seamless integration of LLMs into existing forensic software and investigative workflows. Embedding interactive AI assistants within forensic suites, enabling natural language querying, automated artifact parsing, and AI-driven script generation, can significantly enhance investigative efficiency. Ensuring these integrations are robust and error resistant, and maintaining compatibility with existing forensic processes, evidence documentation systems, and investigative protocols, represents a significant technical challenge. Interdisciplinary collaboration will be crucial in developing user-centric, reliable forensic tools augmented by AI capabilities.

\subsection{Standardized Evaluation and Benchmarking}

A critical gap in current research is the lack of standardized evaluation frameworks for assessing LLM effectiveness and reliability in forensic contexts. Developing shared benchmark datasets, standardized metrics for accuracy, explainability, and utility, and consistent evaluation methodologies is essential for objectively comparing different LLM approaches. Community-driven benchmarking initiatives, similar to established cybersecurity and computer vision evaluations, should be prioritized to accelerate progress and ensure rigorous validation of AI-assisted forensic tools.

\section{Conclusion}
Large Language Models (LLMs) have emerged as transformative tools that significantly automate and augment forensic capabilities, thus reshaping the landscape of digital investigations. This paper systematically explored how LLMs have revolutionized digital forensic approaches, providing a comprehensive and accessible overview for practitioners and researchers alike. Through practical examples and real-world scenarios, we illustrated the superior capabilities of LLMs in enhancing analytical accuracy, efficiency, and scalability in forensic workflows. However, the integration of LLMs into digital forensic processes is not without challenges; issues such as model hallucinations, interpretability, biases, and ethical considerations necessitate cautious and informed application. Addressing these challenges requires further research that focuses on improving transparency, accountability, and standardization in the forensic use of LLM technologies. Ultimately, the thoughtful integration of LLMs holds significant promise in advancing digital forensic practices, fostering trust and reliability, and contributing to more equitable and just judicial outcomes.

\bibliography{typeinst}% common bib file


\begin{thebibliography}{116}
\ifx \bisbn   \undefined \def \bisbn  #1{ISBN #1}\fi
\ifx \binits  \undefined \def \binits#1{#1}\fi
\ifx \bauthor  \undefined \def \bauthor#1{#1}\fi
\ifx \batitle  \undefined \def \batitle#1{#1}\fi
\ifx \bjtitle  \undefined \def \bjtitle#1{#1}\fi
\ifx \bvolume  \undefined \def \bvolume#1{\textbf{#1}}\fi
\ifx \byear  \undefined \def \byear#1{#1}\fi
\ifx \bissue  \undefined \def \bissue#1{#1}\fi
\ifx \bfpage  \undefined \def \bfpage#1{#1}\fi
\ifx \blpage  \undefined \def \blpage #1{#1}\fi
\ifx \burl  \undefined \def \burl#1{\textsf{#1}}\fi
\ifx \doiurl  \undefined \def \doiurl#1{\url{https://doi.org/#1}}\fi
\ifx \betal  \undefined \def \betal{\textit{et al.}}\fi
\ifx \binstitute  \undefined \def \binstitute#1{#1}\fi
\ifx \binstitutionaled  \undefined \def \binstitutionaled#1{#1}\fi
\ifx \bctitle  \undefined \def \bctitle#1{#1}\fi
\ifx \beditor  \undefined \def \beditor#1{#1}\fi
\ifx \bpublisher  \undefined \def \bpublisher#1{#1}\fi
\ifx \bbtitle  \undefined \def \bbtitle#1{#1}\fi
\ifx \bedition  \undefined \def \bedition#1{#1}\fi
\ifx \bseriesno  \undefined \def \bseriesno#1{#1}\fi
\ifx \blocation  \undefined \def \blocation#1{#1}\fi
\ifx \bsertitle  \undefined \def \bsertitle#1{#1}\fi
\ifx \bsnm \undefined \def \bsnm#1{#1}\fi
\ifx \bsuffix \undefined \def \bsuffix#1{#1}\fi
\ifx \bparticle \undefined \def \bparticle#1{#1}\fi
\ifx \barticle \undefined \def \barticle#1{#1}\fi
\bibcommenthead
\ifx \bconfdate \undefined \def \bconfdate #1{#1}\fi
\ifx \botherref \undefined \def \botherref #1{#1}\fi
\ifx \url \undefined \def \url#1{\textsf{#1}}\fi
\ifx \bchapter \undefined \def \bchapter#1{#1}\fi
\ifx \bbook \undefined \def \bbook#1{#1}\fi
\ifx \bcomment \undefined \def \bcomment#1{#1}\fi
\ifx \oauthor \undefined \def \oauthor#1{#1}\fi
\ifx \citeauthoryear \undefined \def \citeauthoryear#1{#1}\fi
\ifx \endbibitem  \undefined \def \endbibitem {}\fi
\ifx \bconflocation  \undefined \def \bconflocation#1{#1}\fi
\ifx \arxivurl  \undefined \def \arxivurl#1{\textsf{#1}}\fi
\csname PreBibitemsHook\endcsname

%%% 1
\bibitem[\protect\citeauthoryear{Wickramasekara
  et~al.}{2025}]{wickramasekara2025exploring}
\begin{barticle}
\bauthor{\bsnm{Wickramasekara}, \binits{A.}},
\bauthor{\bsnm{Breitinger}, \binits{F.}},
\bauthor{\bsnm{Scanlon}, \binits{M.}}:
\batitle{Exploring the potential of large language models for improving digital
  forensic investigation efficiency}.
\bjtitle{Forensic Science International: Digital Investigation}
\bvolume{52},
\bfpage{301859}
(\byear{2025})
\end{barticle}
\endbibitem

%%% 2
\bibitem[\protect\citeauthoryear{Rahman et~al.}{2024}]{rahman2024leveraging}
\begin{botherref}
\oauthor{\bsnm{Rahman}, \binits{M.N.}},
\oauthor{\bsnm{Mohammad}, \binits{T.}},
\oauthor{\bsnm{Virtanen}, \binits{S.}}:
Leveraging large language models for network traffic analysis: Design,
  implementation, and evaluation of an llm-powered system for cyber incident
  reconstruction
(2024)
\end{botherref}
\endbibitem

%%% 3
\bibitem[\protect\citeauthoryear{Xu et~al.}{2024}]{xu2024transforming}
\begin{bchapter}
\bauthor{\bsnm{Xu}, \binits{E.}},
\bauthor{\bsnm{Zhang}, \binits{W.}},
\bauthor{\bsnm{Xu}, \binits{W.}}:
\bctitle{Transforming digital forensics with large language models: Unlocking
  automation, insights, and justice}.
In: \bbtitle{Proceedings of the 33rd ACM International Conference on
  Information and Knowledge Management},
pp. \bfpage{5543}--\blpage{5546}
(\byear{2024})
\end{bchapter}
\endbibitem

%%% 4
\bibitem[\protect\citeauthoryear{Rogers}{2006}]{rogers2006two}
\begin{barticle}
\bauthor{\bsnm{Rogers}, \binits{M.K.}}:
\batitle{A two-dimensional circumplex approach to the development of a hacker
  taxonomy}.
\bjtitle{Digital investigation}
\bvolume{3}(\bissue{2}),
\bfpage{97}--\blpage{102}
(\byear{2006})
\end{barticle}
\endbibitem

%%% 5
\bibitem[\protect\citeauthoryear{Ismail}{2017}]{ismail2017sony}
\begin{botherref}
\oauthor{\bsnm{Ismail}, \binits{M.}}:
Sony pictures and the us federal government: a case study analysis of the sony
  pictures entertainment hack crisis using normal accidents theory
(2017)
\end{botherref}
\endbibitem

%%% 6
\bibitem[\protect\citeauthoryear{Marmura and
  Marmura}{2018}]{marmura2018wikileaks}
\begin{botherref}
\oauthor{\bsnm{Marmura}, \binits{S.M.}},
\oauthor{\bsnm{Marmura}, \binits{S.M.}}:
Wikileaks’ american moment: The dnc emails, russiagate and beyond.
The WikiLeaks Paradigm: Paradoxes and Revelations,
109--133
(2018)
\end{botherref}
\endbibitem

%%% 7
\bibitem[\protect\citeauthoryear{Confessore
  et~al.}{2016}]{confessore2016hacked}
\begin{botherref}
\oauthor{\bsnm{Confessore}, \binits{N.}},
\oauthor{\bsnm{Eder}, \binits{S.}},
\oauthor{\bsnm{October}, \binits{L.}}:
In hacked dnc emails, a glimpse of how big money works.
The New York Times
(2016)
\end{botherref}
\endbibitem

%%% 8
\bibitem[\protect\citeauthoryear{Minnaar}{2017}]{minnaar2017online}
\begin{barticle}
\bauthor{\bsnm{Minnaar}, \binits{A.}}:
\batitle{Online'underground'marketplaces for illicit drugs: the prototype case
  of the dark web website'silk road}.
\bjtitle{Acta Criminologica: African Journal of Criminology \& Victimology}
\bvolume{30}(\bissue{1}),
\bfpage{23}--\blpage{47}
(\byear{2017})
\end{barticle}
\endbibitem

%%% 9
\bibitem[\protect\citeauthoryear{Lacson and Jones}{2016}]{lacson201621st}
\begin{barticle}
\bauthor{\bsnm{Lacson}, \binits{W.}},
\bauthor{\bsnm{Jones}, \binits{B.}}:
\batitle{The 21st century darknet market: lessons from the fall of silk road}.
\bjtitle{International Journal of Cyber Criminology}
\bvolume{10}(\bissue{1}),
\bfpage{40}
(\byear{2016})
\end{barticle}
\endbibitem

%%% 10
\bibitem[\protect\citeauthoryear{Negangard and
  Fay}{2020}]{negangard2020electronic}
\begin{barticle}
\bauthor{\bsnm{Negangard}, \binits{E.M.}},
\bauthor{\bsnm{Fay}, \binits{R.G.}}:
\batitle{Electronic discovery (ediscovery): Performing the early stages of the
  enron investigation}.
\bjtitle{Issues in Accounting Education}
\bvolume{35}(\bissue{1}),
\bfpage{43}--\blpage{58}
(\byear{2020})
\end{barticle}
\endbibitem

%%% 11
\bibitem[\protect\citeauthoryear{Kim et~al.}{}]{kim5110258digital}
\begin{botherref}
\oauthor{\bsnm{Kim}, \binits{K.}},
\oauthor{\bsnm{Lee}, \binits{C.}},
\oauthor{\bsnm{Bae}, \binits{S.}},
\oauthor{\bsnm{Choi}, \binits{J.}},
\oauthor{\bsnm{Kang}, \binits{W.}}:
Digital forensics in law enforcement: A case study of llm-driven evidence
  analysis.
Available at SSRN 5110258
\end{botherref}
\endbibitem

%%% 12
\bibitem[\protect\citeauthoryear{Quick and Choo}{2018}]{quick2018digital}
\begin{barticle}
\bauthor{\bsnm{Quick}, \binits{D.}},
\bauthor{\bsnm{Choo}, \binits{K.-K.R.}}:
\batitle{Digital forensic intelligence: Data subsets and open source
  intelligence (dfint+ osint): A timely and cohesive mix}.
\bjtitle{Future Generation Computer Systems}
\bvolume{78},
\bfpage{558}--\blpage{567}
(\byear{2018})
\end{barticle}
\endbibitem

%%% 13
\bibitem[\protect\citeauthoryear{Chen}{2014}]{chen2014cloud}
\begin{bchapter}
\bauthor{\bsnm{Chen}, \binits{H.-Y.}}:
\bctitle{Cloud crime to traditional digital forensic legal and technical
  challenges and countermeasures}.
In: \bbtitle{2014 IEEE Workshop on Advanced Research and Technology in Industry
  Applications (WARTIA)},
pp. \bfpage{990}--\blpage{994}
(\byear{2014}).
\bcomment{IEEE}
\end{bchapter}
\endbibitem

%%% 14
\bibitem[\protect\citeauthoryear{Fernando}{2023}]{fernando2023multidimensional}
\begin{barticle}
\bauthor{\bsnm{Fernando}, \binits{K.}}:
\batitle{A multidimensional framework for utilizing big data analytics and ai
  in strengthening digital forensics and cybersecurity investigations}.
\bjtitle{International Journal of Cybersecurity Risk Management, Forensics, and
  Compliance}
\bvolume{7}(\bissue{12}),
\bfpage{16}--\blpage{30}
(\byear{2023})
\end{barticle}
\endbibitem

%%% 15
\bibitem[\protect\citeauthoryear{Malik et~al.}{2024}]{malik2024cloud}
\begin{barticle}
\bauthor{\bsnm{Malik}, \binits{A.W.}},
\bauthor{\bsnm{Bhatti}, \binits{D.S.}},
\bauthor{\bsnm{Park}, \binits{T.-J.}},
\bauthor{\bsnm{Ishtiaq}, \binits{H.U.}},
\bauthor{\bsnm{Ryou}, \binits{J.-C.}},
\bauthor{\bsnm{Kim}, \binits{K.-I.}}:
\batitle{Cloud digital forensics: Beyond tools, techniques, and challenges}.
\bjtitle{Sensors}
\bvolume{24}(\bissue{2}),
\bfpage{433}
(\byear{2024})
\end{barticle}
\endbibitem

%%% 16
\bibitem[\protect\citeauthoryear{Garach et~al.}{2024}]{garach2024comprehensive}
\begin{botherref}
\oauthor{\bsnm{Garach}, \binits{J.}},
\oauthor{\bsnm{Singh}, \binits{S.K.}},
\oauthor{\bsnm{Reddy}, \binits{A.P.C.}},
\oauthor{\bsnm{Khan}, \binits{H.}}, et al.:
A comprehensive review on artificial intelligence in digital forensics with
  taxonomies, issues, and solutions: Ai in digital forensics.
Strategies for E-Commerce Data Security: Cloud, Blockchain, AI, and Machine
  Learning,
1--28
(2024)
\end{botherref}
\endbibitem

%%% 17
\bibitem[\protect\citeauthoryear{Wang et~al.}{2023}]{wang2023preventing}
\begin{bchapter}
\bauthor{\bsnm{Wang}, \binits{Z.}},
\bauthor{\bsnm{Saxena}, \binits{N.}},
\bauthor{\bsnm{Yu}, \binits{T.}},
\bauthor{\bsnm{Karki}, \binits{S.}},
\bauthor{\bsnm{Zetty}, \binits{T.}},
\bauthor{\bsnm{Haque}, \binits{I.}},
\bauthor{\bsnm{Zhou}, \binits{S.}},
\bauthor{\bsnm{Kc}, \binits{D.}},
\bauthor{\bsnm{Stockwell}, \binits{I.}},
\bauthor{\bsnm{Bifet}, \binits{A.}}, \betal:
\bctitle{Preventing discriminatory decision-making in evolving data streams}.
In: \bbtitle{Proceedings of the 2023 ACM Conference on Fairness,
  Accountability, and Transparency (FAccT)}
(\byear{2023})
\end{bchapter}
\endbibitem

%%% 18
\bibitem[\protect\citeauthoryear{Zhang et~al.}{2023}]{zhang2023individual}
\begin{bchapter}
\bauthor{\bsnm{Zhang}, \binits{W.}},
\bauthor{\bsnm{Wang}, \binits{Z.}},
\bauthor{\bsnm{Kim}, \binits{J.}},
\bauthor{\bsnm{Cheng}, \binits{C.}},
\bauthor{\bsnm{Oommen}, \binits{T.}},
\bauthor{\bsnm{Ravikumar}, \binits{P.}},
\bauthor{\bsnm{Weiss}, \binits{J.}}:
\bctitle{Individual fairness under uncertainty}.
In: \bbtitle{26th European Conference on Artificial Intelligence},
pp. \bfpage{3042}--\blpage{3049}
(\byear{2023})
\end{bchapter}
\endbibitem

%%% 19
\bibitem[\protect\citeauthoryear{Wang et~al.}{2023}]{wang2023fg2an}
\begin{bchapter}
\bauthor{\bsnm{Wang}, \binits{Z.}},
\bauthor{\bsnm{Wallace}, \binits{C.}},
\bauthor{\bsnm{Bifet}, \binits{A.}},
\bauthor{\bsnm{Yao}, \binits{X.}},
\bauthor{\bsnm{Zhang}, \binits{W.}}:
\bctitle{Fg$^2$an: Fairness-aware graph generative adversarial networks}.
In: \bbtitle{Joint European Conference on Machine Learning and Knowledge
  Discovery in Databases},
pp. \bfpage{259}--\blpage{275}
(\byear{2023}).
\bcomment{Springer Nature Switzerland}
\end{bchapter}
\endbibitem

%%% 20
\bibitem[\protect\citeauthoryear{Yazdani
  et~al.}{2024}]{yazdani2024comprehensive}
\begin{botherref}
\oauthor{\bsnm{Yazdani}, \binits{S.}},
\oauthor{\bsnm{Saxena}, \binits{N.}},
\oauthor{\bsnm{Wang}, \binits{Z.}},
\oauthor{\bsnm{Wu}, \binits{Y.}},
\oauthor{\bsnm{Zhang}, \binits{W.}}:
A comprehensive survey of image and video generative ai: Recent advances,
  variants, and applications
(2024)
\end{botherref}
\endbibitem

%%% 21
\bibitem[\protect\citeauthoryear{Wang et~al.}{2023}]{wang2023mitigating}
\begin{bchapter}
\bauthor{\bsnm{Wang}, \binits{Z.}},
\bauthor{\bsnm{Narasimhan}, \binits{G.}},
\bauthor{\bsnm{Yao}, \binits{X.}},
\bauthor{\bsnm{Zhang}, \binits{W.}}:
\bctitle{Mitigating multisource biases in graph neural networks via real
  counterfactual samples}.
In: \bbtitle{2023 IEEE International Conference on Data Mining (ICDM)},
pp. \bfpage{638}--\blpage{647}
(\byear{2023}).
\bcomment{IEEE}
\end{bchapter}
\endbibitem

%%% 22
\bibitem[\protect\citeauthoryear{Chinta et~al.}{2023}]{chinta2023optimization}
\begin{bchapter}
\bauthor{\bsnm{Chinta}, \binits{S.V.}},
\bauthor{\bsnm{Fernandes}, \binits{K.}},
\bauthor{\bsnm{Cheng}, \binits{N.}},
\bauthor{\bsnm{Fernandez}, \binits{J.}},
\bauthor{\bsnm{Yazdani}, \binits{S.}},
\bauthor{\bsnm{Yin}, \binits{Z.}},
\bauthor{\bsnm{Wang}, \binits{Z.}},
\bauthor{\bsnm{Wang}, \binits{X.}},
\bauthor{\bsnm{Xu}, \binits{W.}},
\bauthor{\bsnm{Liu}, \binits{J.}}, \betal:
\bctitle{Optimization and improvement of fake news detection using voting
  technique for societal benefit}.
In: \bbtitle{2023 IEEE International Conference on Data Mining Workshops
  (ICDMW)},
pp. \bfpage{1565}--\blpage{1574}
(\byear{2023}).
\bcomment{IEEE}
\end{bchapter}
\endbibitem

%%% 23
\bibitem[\protect\citeauthoryear{Wang et~al.}{2024}]{wang2024history}
\begin{botherref}
\oauthor{\bsnm{Wang}, \binits{Z.}},
\oauthor{\bsnm{Chu}, \binits{Z.}},
\oauthor{\bsnm{Doan}, \binits{T.V.}},
\oauthor{\bsnm{Ni}, \binits{S.}},
\oauthor{\bsnm{Yang}, \binits{M.}},
\oauthor{\bsnm{Zhang}, \binits{W.}}:
History, development, and principles of large language models: an introductory
  survey.
AI and Ethics,
1--17
(2024)
\end{botherref}
\endbibitem

%%% 24
\bibitem[\protect\citeauthoryear{Chu et~al.}{2024}]{chu2024fairness}
\begin{barticle}
\bauthor{\bsnm{Chu}, \binits{Z.}},
\bauthor{\bsnm{Wang}, \binits{Z.}},
\bauthor{\bsnm{Zhang}, \binits{W.}}:
\batitle{Fairness in large language models: A taxonomic survey}.
\bjtitle{ACM SIGKDD explorations newsletter}
\bvolume{26}(\bissue{1}),
\bfpage{34}--\blpage{48}
(\byear{2024})
\end{barticle}
\endbibitem

%%% 25
\bibitem[\protect\citeauthoryear{Dzuong et~al.}{2024}]{dzuong2024uncertain}
\begin{botherref}
\oauthor{\bsnm{Dzuong}, \binits{J.}},
\oauthor{\bsnm{Wang}, \binits{Z.}},
\oauthor{\bsnm{Zhang}, \binits{W.}}:
Uncertain boundaries: Multidisciplinary approaches to copyright issues in
  generative ai.
arXiv preprint arXiv:2404.08221
(2024)
\end{botherref}
\endbibitem

%%% 26
\bibitem[\protect\citeauthoryear{Yin et~al.}{2024}]{yin2024improving}
\begin{bchapter}
\bauthor{\bsnm{Yin}, \binits{Z.}},
\bauthor{\bsnm{Wang}, \binits{Z.}},
\bauthor{\bsnm{Zhang}, \binits{W.}}:
\bctitle{Improving fairness in machine learning software via counterfactual
  fairness thinking}.
In: \bbtitle{Proceedings of the 2024 IEEE/ACM 46th International Conference on
  Software Engineering: Companion Proceedings},
pp. \bfpage{420}--\blpage{421}
(\byear{2024})
\end{bchapter}
\endbibitem

%%% 27
\bibitem[\protect\citeauthoryear{Wang et~al.}{2023}]{wang2023towards}
\begin{botherref}
\oauthor{\bsnm{Wang}, \binits{Z.}},
\oauthor{\bsnm{Zhou}, \binits{Y.}},
\oauthor{\bsnm{Haque}, \binits{I.}},
\oauthor{\bsnm{Lo}, \binits{D.}},
\oauthor{\bsnm{Zhang}, \binits{W.}}:
Towards fair machine learning software: Understanding and addressing model bias
  through counterfactual thinking.
arXiv preprint arXiv:2302.08018
(2023)
\end{botherref}
\endbibitem

%%% 28
\bibitem[\protect\citeauthoryear{Wang et~al.}{2024}]{wang2024toward}
\begin{botherref}
\oauthor{\bsnm{Wang}, \binits{Z.}},
\oauthor{\bsnm{Qiu}, \binits{M.}},
\oauthor{\bsnm{Chen}, \binits{M.}},
\oauthor{\bsnm{Salem}, \binits{M.B.}},
\oauthor{\bsnm{Yao}, \binits{X.}},
\oauthor{\bsnm{Zhang}, \binits{W.}}:
Toward fair graph neural networks via real counterfactual samples.
Knowledge and Information Systems,
1--25
(2024)
\end{botherref}
\endbibitem

%%% 29
\bibitem[\protect\citeauthoryear{Chinta et~al.}{2024}]{chinta2024fairaied}
\begin{botherref}
\oauthor{\bsnm{Chinta}, \binits{S.V.}},
\oauthor{\bsnm{Wang}, \binits{Z.}},
\oauthor{\bsnm{Yin}, \binits{Z.}},
\oauthor{\bsnm{Hoang}, \binits{N.}},
\oauthor{\bsnm{Gonzalez}, \binits{M.}},
\oauthor{\bsnm{Quy}, \binits{T.L.}},
\oauthor{\bsnm{Zhang}, \binits{W.}}:
Fairaied: Navigating fairness, bias, and ethics in educational ai applications.
arXiv preprint arXiv:2407.18745
(2024)
\end{botherref}
\endbibitem

%%% 30
\bibitem[\protect\citeauthoryear{Doan et~al.}{2024}]{doan2024fairness}
\begin{bchapter}
\bauthor{\bsnm{Doan}, \binits{T.V.}},
\bauthor{\bsnm{Wang}, \binits{Z.}},
\bauthor{\bsnm{Hoang}, \binits{N.N.M.}},
\bauthor{\bsnm{Zhang}, \binits{W.}}:
\bctitle{Fairness in large language models in three hours}.
In: \bbtitle{Proceedings of the 33rd ACM International Conference on
  Information and Knowledge Management},
pp. \bfpage{5514}--\blpage{5517}
(\byear{2024})
\end{bchapter}
\endbibitem

%%% 31
\bibitem[\protect\citeauthoryear{Chinta et~al.}{2024}]{chinta2024ai}
\begin{botherref}
\oauthor{\bsnm{Chinta}, \binits{S.V.}},
\oauthor{\bsnm{Wang}, \binits{Z.}},
\oauthor{\bsnm{Zhang}, \binits{X.}},
\oauthor{\bsnm{Viet}, \binits{T.D.}},
\oauthor{\bsnm{Kashif}, \binits{A.}},
\oauthor{\bsnm{Smith}, \binits{M.A.}},
\oauthor{\bsnm{Zhang}, \binits{W.}}:
Ai-driven healthcare: A survey on ensuring fairness and mitigating bias.
arXiv preprint arXiv:2407.19655
(2024)
\end{botherref}
\endbibitem

%%% 32
\bibitem[\protect\citeauthoryear{Wang et~al.}{2024a}]{wang2024individual1}
\begin{bchapter}
\bauthor{\bsnm{Wang}, \binits{Z.}},
\bauthor{\bsnm{Dzuong}, \binits{J.}},
\bauthor{\bsnm{Yuan}, \binits{X.}},
\bauthor{\bsnm{Chen}, \binits{Z.}},
\bauthor{\bsnm{Wu}, \binits{Y.}},
\bauthor{\bsnm{Yao}, \binits{X.}},
\bauthor{\bsnm{Zhang}, \binits{W.}}:
\bctitle{Individual fairness with group awareness under uncertainty}.
In: \bbtitle{Joint European Conference on Machine Learning and Knowledge
  Discovery in Databases},
pp. \bfpage{89}--\blpage{106}
(\byear{2024}).
\bcomment{Springer Nature Switzerland}
\end{bchapter}
\endbibitem

%%% 33
\bibitem[\protect\citeauthoryear{Wang et~al.}{2024b}]{doan2024fairness1}
\begin{botherref}
\oauthor{\bsnm{Wang}, \binits{Z.}},
\oauthor{\bsnm{Palikhe}, \binits{A.}},
\oauthor{\bsnm{Yin}, \binits{Z.}},
\oauthor{\bsnm{Zhang}, \binits{W.}}:
Fairness definitions in language models explained.
arXiv preprint arXiv:2407.18454
(2024)
\end{botherref}
\endbibitem

%%% 34
\bibitem[\protect\citeauthoryear{Wang et~al.}{2024c}]{wang2024advancing}
\begin{bchapter}
\bauthor{\bsnm{Wang}, \binits{Z.}},
\bauthor{\bsnm{Chu}, \binits{Z.}},
\bauthor{\bsnm{Blanco}, \binits{R.}},
\bauthor{\bsnm{Chen}, \binits{Z.}},
\bauthor{\bsnm{Chen}, \binits{S.-C.}},
\bauthor{\bsnm{Zhang}, \binits{W.}}:
\bctitle{Advancing graph counterfactual fairness through fair representation
  learning}.
In: \bbtitle{Joint European Conference on Machine Learning and Knowledge
  Discovery in Databases},
pp. \bfpage{40}--\blpage{58}
(\byear{2024}).
\bcomment{Springer Nature Switzerland}
\end{bchapter}
\endbibitem

%%% 35
\bibitem[\protect\citeauthoryear{Swanson}{}]{swansonbullets}
\begin{botherref}
\oauthor{\bsnm{Swanson}, \binits{C.}}:
Bullets and ballots: Exploring the effects of nearly successful assassination
  attempts on general election performance in the united states.
The UWJPS is thankful for the continued support of the Department of Political
  Science at the University of Washington. In addition, we are grateful to the
  students who submitted their work and ideas.,
15
\end{botherref}
\endbibitem

%%% 36
\bibitem[\protect\citeauthoryear{{The New York
  Times}}{2024}]{nyt_trump_shooting_2024}
\begin{botherref}
\oauthor{\bsnm{{The New York Times}}}:
Investigators Unlock Gunman’s Phone in Search for Motive in Trump Shooting.
Accessed: 2025-03-30.
\url{https://www.nytimes.com/live/2024/07/15/us/trump-shooting-investigation}
\end{botherref}
\endbibitem

%%% 37
\bibitem[\protect\citeauthoryear{Bartoletti
  et~al.}{2021}]{bartoletti2021cryptocurrency}
\begin{barticle}
\bauthor{\bsnm{Bartoletti}, \binits{M.}},
\bauthor{\bsnm{Lande}, \binits{S.}},
\bauthor{\bsnm{Loddo}, \binits{A.}},
\bauthor{\bsnm{Pompianu}, \binits{L.}},
\bauthor{\bsnm{Serusi}, \binits{S.}}:
\batitle{Cryptocurrency scams: analysis and perspectives}.
\bjtitle{Ieee Access}
\bvolume{9},
\bfpage{148353}--\blpage{148373}
(\byear{2021})
\end{barticle}
\endbibitem

%%% 38
\bibitem[\protect\citeauthoryear{Cimpanu}{2020}]{cimpanu2020fbi}
\begin{botherref}
\oauthor{\bsnm{Cimpanu}, \binits{C.}}:
How the FBI Tracked down the Twitter Hackers.
Accessed: 2025-03-30.
\url{https://www.zdnet.com/article/how-the-fbi-tracked-down-the-twitter-hackers/}
\end{botherref}
\endbibitem

%%% 39
\bibitem[\protect\citeauthoryear{Kessler and
  Phillips}{2020}]{kessler2020cryptography}
\begin{barticle}
\bauthor{\bsnm{Kessler}, \binits{G.C.}},
\bauthor{\bsnm{Phillips}, \binits{A.M.}}:
\batitle{Cryptography, passwords, privacy, and the fifth amendment}.
\bjtitle{Journal of Digital Forensics, Security and Law}
\bvolume{15}(\bissue{2}),
\bfpage{2}
(\byear{2020})
\end{barticle}
\endbibitem

%%% 40
\bibitem[\protect\citeauthoryear{Clarke}{2020}]{clarke2020pensacola}
\begin{botherref}
\oauthor{\bsnm{Clarke}, \binits{C.}}:
The pensacola terrorist attack: The enduring influence of al-qaida and its
  affiliates.
CTC Sentinel
\textbf{13}(3)
(2020)
\end{botherref}
\endbibitem

%%% 41
\bibitem[\protect\citeauthoryear{Vaghela et~al.}{2024}]{vaghela2024digital}
\begin{bchapter}
\bauthor{\bsnm{Vaghela}, \binits{R.}},
\bauthor{\bsnm{Gowda}, \binits{V.D.}},
\bauthor{\bsnm{Taj}, \binits{M.}},
\bauthor{\bsnm{Arudra}, \binits{A.}},
\bauthor{\bsnm{Chopra}, \binits{M.}}:
\bctitle{Digital evidence collection and preservation in computer network
  forensics}.
In: \bbtitle{Handbook of Research on Innovative Approaches to Information
  Technology in Library and Information Science},
pp. \bfpage{42}--\blpage{62}
(\byear{2024})
\end{bchapter}
\endbibitem

%%% 42
\bibitem[\protect\citeauthoryear{Allam}{2020}]{allam2020pensacola}
\begin{botherref}
\oauthor{\bsnm{Allam}, \binits{H.}}:
{FBI: New iPhone Evidence Shows Pensacola Shooter Had Ties To Al-Qaida}.
Accessed: 2025-03-30.
\url{https://www.npr.org/2020/05/18/857932909/fbi-new-iphone-evidence-shows-pensacola-shooter-had-ties-to-al-qaida}
\end{botherref}
\endbibitem

%%% 43
\bibitem[\protect\citeauthoryear{Nayak}{2024}]{nayak2024ai}
\begin{barticle}
\bauthor{\bsnm{Nayak}, \binits{M.}}:
\batitle{Ai-enhanced digital forensics: Automated techniques for efficient
  investigation and evidence collection}.
\bjtitle{J. Electrical Systems}
\bvolume{20}(\bissue{1s}),
\bfpage{211}--\blpage{229}
(\byear{2024})
\end{barticle}
\endbibitem

%%% 44
\bibitem[\protect\citeauthoryear{Akeiber}{2025}]{akeiber2025comprehensive}
\begin{botherref}
\oauthor{\bsnm{Akeiber}, \binits{H.J.}}:
A comprehensive study of cybercrime and digital forensics through machine
  learning and ai.
Al-Rafidain Journal of Engineering Sciences,
369--395
(2025)
\end{botherref}
\endbibitem

%%% 45
\bibitem[\protect\citeauthoryear{Liu et~al.}{2024}]{liu2024toward}
\begin{botherref}
\oauthor{\bsnm{Liu}, \binits{J.}},
\oauthor{\bsnm{Kong}, \binits{Z.}},
\oauthor{\bsnm{Zhao}, \binits{P.}},
\oauthor{\bsnm{Yang}, \binits{C.}},
\oauthor{\bsnm{Tang}, \binits{H.}},
\oauthor{\bsnm{Shen}, \binits{X.}},
\oauthor{\bsnm{Yuan}, \binits{G.}},
\oauthor{\bsnm{Niu}, \binits{W.}},
\oauthor{\bsnm{Zhang}, \binits{W.}},
\oauthor{\bsnm{Lin}, \binits{X.}}, et al.:
Toward adaptive large language models structured pruning via hybrid-grained
  weight importance assessment.
arXiv preprint arXiv:2403.10799
(2024)
\end{botherref}
\endbibitem

%%% 46
\bibitem[\protect\citeauthoryear{Jin et~al.}{2023}]{jin2023rethinking}
\begin{bchapter}
\bauthor{\bsnm{Jin}, \binits{H.}},
\bauthor{\bsnm{Wei}, \binits{W.}},
\bauthor{\bsnm{Wang}, \binits{X.}},
\bauthor{\bsnm{Zhang}, \binits{W.}},
\bauthor{\bsnm{Wu}, \binits{Y.}}:
\bctitle{Rethinking learning rate tuning in the era of large language models}.
In: \bbtitle{2023 IEEE 5th International Conference on Cognitive Machine
  Intelligence (CogMI)},
pp. \bfpage{112}--\blpage{121}
(\byear{2023}).
\bcomment{IEEE}
\end{bchapter}
\endbibitem

%%% 47
\bibitem[\protect\citeauthoryear{Ferrag et~al.}{2025}]{ferrag2025generative}
\begin{botherref}
\oauthor{\bsnm{Ferrag}, \binits{M.A.}},
\oauthor{\bsnm{Alwahedi}, \binits{F.}},
\oauthor{\bsnm{Battah}, \binits{A.}},
\oauthor{\bsnm{Cherif}, \binits{B.}},
\oauthor{\bsnm{Mechri}, \binits{A.}},
\oauthor{\bsnm{Tihanyi}, \binits{N.}},
\oauthor{\bsnm{Bisztray}, \binits{T.}},
\oauthor{\bsnm{Debbah}, \binits{M.}}:
Generative ai in cybersecurity: A comprehensive review of llm applications and
  vulnerabilities.
Internet of Things and Cyber-Physical Systems
(2025)
\end{botherref}
\endbibitem

%%% 48
\bibitem[\protect\citeauthoryear{Yao et~al.}{2024}]{yao2024survey}
\begin{botherref}
\oauthor{\bsnm{Yao}, \binits{Y.}},
\oauthor{\bsnm{Duan}, \binits{J.}},
\oauthor{\bsnm{Xu}, \binits{K.}},
\oauthor{\bsnm{Cai}, \binits{Y.}},
\oauthor{\bsnm{Sun}, \binits{Z.}},
\oauthor{\bsnm{Zhang}, \binits{Y.}}:
A survey on large language model (llm) security and privacy: The good, the bad,
  and the ugly.
High-Confidence Computing,
100211
(2024)
\end{botherref}
\endbibitem

%%% 49
\bibitem[\protect\citeauthoryear{Valmeekam et~al.}{2022}]{valmeekam2022large}
\begin{bchapter}
\bauthor{\bsnm{Valmeekam}, \binits{K.}},
\bauthor{\bsnm{Olmo}, \binits{A.}},
\bauthor{\bsnm{Sreedharan}, \binits{S.}},
\bauthor{\bsnm{Kambhampati}, \binits{S.}}:
\bctitle{Large language models still can't plan (a benchmark for llms on
  planning and reasoning about change)}.
In: \bbtitle{NeurIPS 2022 Foundation Models for Decision Making Workshop}
(\byear{2022})
\end{bchapter}
\endbibitem

%%% 50
\bibitem[\protect\citeauthoryear{Kumarage et~al.}{2024}]{kumarage2024survey}
\begin{botherref}
\oauthor{\bsnm{Kumarage}, \binits{T.}},
\oauthor{\bsnm{Agrawal}, \binits{G.}},
\oauthor{\bsnm{Sheth}, \binits{P.}},
\oauthor{\bsnm{Moraffah}, \binits{R.}},
\oauthor{\bsnm{Chadha}, \binits{A.}},
\oauthor{\bsnm{Garland}, \binits{J.}},
\oauthor{\bsnm{Liu}, \binits{H.}}:
A survey of ai-generated text forensic systems: Detection, attribution, and
  characterization.
arXiv preprint arXiv:2403.01152
(2024)
\end{botherref}
\endbibitem

%%% 51
\bibitem[\protect\citeauthoryear{Ahmed et~al.}{2016}]{ahmed2016survey}
\begin{barticle}
\bauthor{\bsnm{Ahmed}, \binits{M.}},
\bauthor{\bsnm{Mahmood}, \binits{A.N.}},
\bauthor{\bsnm{Hu}, \binits{J.}}:
\batitle{A survey of network anomaly detection techniques}.
\bjtitle{Journal of Network and Computer Applications}
\bvolume{60},
\bfpage{19}--\blpage{31}
(\byear{2016})
\end{barticle}
\endbibitem

%%% 52
\bibitem[\protect\citeauthoryear{Liu et~al.}{2024}]{liu2024large}
\begin{botherref}
\oauthor{\bsnm{Liu}, \binits{C.}},
\oauthor{\bsnm{Xie}, \binits{X.}},
\oauthor{\bsnm{Zhang}, \binits{X.}},
\oauthor{\bsnm{Cui}, \binits{Y.}}:
Large language models for networking: Workflow, advances and challenges.
IEEE Network
(2024)
\end{botherref}
\endbibitem

%%% 53
\bibitem[\protect\citeauthoryear{Velasco}{2022}]{velasco2022cybercrime}
\begin{bchapter}
\bauthor{\bsnm{Velasco}, \binits{C.}}:
\bctitle{Cybercrime and artificial intelligence. an overview of the work of
  international organizations on criminal justice and the international
  applicable instruments}.
In: \bbtitle{ERA Forum},
vol. \bseriesno{23},
pp. \bfpage{109}--\blpage{126}
(\byear{2022}).
\bcomment{Springer}
\end{bchapter}
\endbibitem

%%% 54
\bibitem[\protect\citeauthoryear{Mijwil et~al.}{2023}]{mijwil2023towards}
\begin{barticle}
\bauthor{\bsnm{Mijwil}, \binits{M.M.}},
\bauthor{\bsnm{Aljanabi}, \binits{M.}},
\bauthor{\bsnm{ChatGPT}, \binits{C.}}:
\batitle{Towards artificial intelligence-based cybersecurity: The practices and
  chatgpt generated ways to combat cybercrime}.
\bjtitle{Iraqi Journal For Computer Science and Mathematics}
\bvolume{4}(\bissue{1}),
\bfpage{8}
(\byear{2023})
\end{barticle}
\endbibitem

%%% 55
\bibitem[\protect\citeauthoryear{Zhang and Xie}{2023}]{zhang2023forensiq}
\begin{bchapter}
\bauthor{\bsnm{Zhang}, \binits{R.}},
\bauthor{\bsnm{Xie}, \binits{M.}}:
\bctitle{Forensiq: A knowledge graph question answering system for iot
  forensics}.
In: \bbtitle{International Conference on Digital Forensics and Cyber Crime},
pp. \bfpage{300}--\blpage{314}
(\byear{2023}).
\bcomment{Springer}
\end{bchapter}
\endbibitem

%%% 56
\bibitem[\protect\citeauthoryear{Siddiqui
  et~al.}{2018}]{siddiqui2018application}
\begin{barticle}
\bauthor{\bsnm{Siddiqui}, \binits{M.Z.}},
\bauthor{\bsnm{Yadav}, \binits{S.}},
\bauthor{\bsnm{Husain}, \binits{M.S.}}:
\batitle{Application of artificial intelligence in fighting against cyber
  crimes: a review}.
\bjtitle{Int. J. Adv. Res. Comput. Sci}
\bvolume{9}(\bissue{2}),
\bfpage{118}--\blpage{122}
(\byear{2018})
\end{barticle}
\endbibitem

%%% 57
\bibitem[\protect\citeauthoryear{Chen et~al.}{2024}]{chen2024exploring}
\begin{barticle}
\bauthor{\bsnm{Chen}, \binits{Z.}},
\bauthor{\bsnm{Mao}, \binits{H.}},
\bauthor{\bsnm{Li}, \binits{H.}},
\bauthor{\bsnm{Jin}, \binits{W.}},
\bauthor{\bsnm{Wen}, \binits{H.}},
\bauthor{\bsnm{Wei}, \binits{X.}},
\bauthor{\bsnm{Wang}, \binits{S.}},
\bauthor{\bsnm{Yin}, \binits{D.}},
\bauthor{\bsnm{Fan}, \binits{W.}},
\bauthor{\bsnm{Liu}, \binits{H.}}, \betal:
\batitle{Exploring the potential of large language models (llms) in learning on
  graphs}.
\bjtitle{ACM SIGKDD Explorations Newsletter}
\bvolume{25}(\bissue{2}),
\bfpage{42}--\blpage{61}
(\byear{2024})
\end{barticle}
\endbibitem

%%% 58
\bibitem[\protect\citeauthoryear{Yang et~al.}{2024}]{yang2024harnessing}
\begin{barticle}
\bauthor{\bsnm{Yang}, \binits{J.}},
\bauthor{\bsnm{Jin}, \binits{H.}},
\bauthor{\bsnm{Tang}, \binits{R.}},
\bauthor{\bsnm{Han}, \binits{X.}},
\bauthor{\bsnm{Feng}, \binits{Q.}},
\bauthor{\bsnm{Jiang}, \binits{H.}},
\bauthor{\bsnm{Zhong}, \binits{S.}},
\bauthor{\bsnm{Yin}, \binits{B.}},
\bauthor{\bsnm{Hu}, \binits{X.}}:
\batitle{Harnessing the power of llms in practice: A survey on chatgpt and
  beyond}.
\bjtitle{ACM Transactions on Knowledge Discovery from Data}
\bvolume{18}(\bissue{6}),
\bfpage{1}--\blpage{32}
(\byear{2024})
\end{barticle}
\endbibitem

%%% 59
\bibitem[\protect\citeauthoryear{Smirnov}{2025}]{smirnov2025enhancing}
\begin{barticle}
\bauthor{\bsnm{Smirnov}, \binits{E.}}:
\batitle{Enhancing qualitative research in psychology with large language
  models: a methodological exploration and examples of simulations}.
\bjtitle{Qualitative Research in Psychology}
\bvolume{22}(\bissue{2}),
\bfpage{482}--\blpage{512}
(\byear{2025})
\end{barticle}
\endbibitem

%%% 60
\bibitem[\protect\citeauthoryear{Kao}{2025}]{kao2025accelerating}
\begin{botherref}
\oauthor{\bsnm{Kao}, \binits{H.-H.}}:
Accelerating multilingual cryptocurrency forensics: An nlp-driven approach for
  efficient mnemonic identification.
IEEE Access
(2025)
\end{botherref}
\endbibitem

%%% 61
\bibitem[\protect\citeauthoryear{Karie et~al.}{2019}]{karie2019diverging}
\begin{barticle}
\bauthor{\bsnm{Karie}, \binits{N.M.}},
\bauthor{\bsnm{Kebande}, \binits{V.R.}},
\bauthor{\bsnm{Venter}, \binits{H.}}:
\batitle{Diverging deep learning cognitive computing techniques into cyber
  forensics}.
\bjtitle{Forensic Science International: Synergy}
\bvolume{1},
\bfpage{61}--\blpage{67}
(\byear{2019})
\end{barticle}
\endbibitem

%%% 62
\bibitem[\protect\citeauthoryear{Arshad et~al.}{2018}]{arshad2018digital}
\begin{barticle}
\bauthor{\bsnm{Arshad}, \binits{H.}},
\bauthor{\bsnm{Jantan}, \binits{A.B.}},
\bauthor{\bsnm{Abiodun}, \binits{O.I.}}:
\batitle{Digital forensics: review of issues in scientific validation of
  digital evidence}.
\bjtitle{Journal of Information Processing Systems}
\bvolume{14}(\bissue{2}),
\bfpage{346}--\blpage{376}
(\byear{2018})
\end{barticle}
\endbibitem

%%% 63
\bibitem[\protect\citeauthoryear{Klas{\'e}n et~al.}{2024}]{klasen2024invisible}
\begin{barticle}
\bauthor{\bsnm{Klas{\'e}n}, \binits{L.}},
\bauthor{\bsnm{Fock}, \binits{N.}},
\bauthor{\bsnm{Forchheimer}, \binits{R.}}:
\batitle{The invisible evidence: Digital forensics as key to solving crimes in
  the digital age}.
\bjtitle{Forensic science international}
\bvolume{362},
\bfpage{112133}
(\byear{2024})
\end{barticle}
\endbibitem

%%% 64
\bibitem[\protect\citeauthoryear{Daniel and Daniel}{2011}]{daniel2011digital}
\begin{bbook}
\bauthor{\bsnm{Daniel}, \binits{L.}},
\bauthor{\bsnm{Daniel}, \binits{L.}}:
\bbtitle{Digital Forensics for Legal Professionals: Understanding Digital
  Evidence from the Warrant to the Courtroom},
(\byear{2011})
\end{bbook}
\endbibitem

%%% 65
\bibitem[\protect\citeauthoryear{Caballero}{2024}]{caballero2024leveraging}
\begin{botherref}
\oauthor{\bsnm{Caballero}, \binits{E.Q.}}:
Leveraging large language models for legal document understanding and software
  system analysis: Addressing key challenges.
PhD thesis,
Baylor University
(2024)
\end{botherref}
\endbibitem

%%% 66
\bibitem[\protect\citeauthoryear{Akhtar et~al.}{2025}]{akhtar2025llm}
\begin{botherref}
\oauthor{\bsnm{Akhtar}, \binits{S.}},
\oauthor{\bsnm{Khan}, \binits{S.}},
\oauthor{\bsnm{Parkinson}, \binits{S.}}:
Llm-based event log analysis techniques: A survey.
arXiv preprint arXiv:2502.00677
(2025)
\end{botherref}
\endbibitem

%%% 67
\bibitem[\protect\citeauthoryear{Labajov{\'a}}{2023}]{labajova2023state}
\begin{botherref}
\oauthor{\bsnm{Labajov{\'a}}, \binits{L.}}:
The state of AI: Exploring the perceptions, credibility, and trustworthiness of
  the users towards AI-Generated Content
(2023)
\end{botherref}
\endbibitem

%%% 68
\bibitem[\protect\citeauthoryear{Khlaif et~al.}{2023}]{khlaif2023potential}
\begin{barticle}
\bauthor{\bsnm{Khlaif}, \binits{Z.N.}},
\bauthor{\bsnm{Mousa}, \binits{A.}},
\bauthor{\bsnm{Hattab}, \binits{M.K.}},
\bauthor{\bsnm{Itmazi}, \binits{J.}},
\bauthor{\bsnm{Hassan}, \binits{A.A.}},
\bauthor{\bsnm{Sanmugam}, \binits{M.}},
\bauthor{\bsnm{Ayyoub}, \binits{A.}}:
\batitle{The potential and concerns of using ai in scientific research: Chatgpt
  performance evaluation}.
\bjtitle{JMIR Medical Education}
\bvolume{9},
\bfpage{47049}
(\byear{2023})
\end{barticle}
\endbibitem

%%% 69
\bibitem[\protect\citeauthoryear{Raza}{2024}]{raza2024ai}
\begin{barticle}
\bauthor{\bsnm{Raza}, \binits{H.}}:
\batitle{Ai-driven assessment: Reliability, bias, and ethical implications}.
\bjtitle{Journal of AI in Education: Innovations, Opportunities, Challenges,
  and Future Directions}
\bvolume{1}(\bissue{2}),
\bfpage{36}--\blpage{47}
(\byear{2024})
\end{barticle}
\endbibitem

%%% 70
\bibitem[\protect\citeauthoryear{Azodi et~al.}{2020}]{azodi2020opening}
\begin{barticle}
\bauthor{\bsnm{Azodi}, \binits{C.B.}},
\bauthor{\bsnm{Tang}, \binits{J.}},
\bauthor{\bsnm{Shiu}, \binits{S.-H.}}:
\batitle{Opening the black box: interpretable machine learning for
  geneticists}.
\bjtitle{Trends in genetics}
\bvolume{36}(\bissue{6}),
\bfpage{442}--\blpage{455}
(\byear{2020})
\end{barticle}
\endbibitem

%%% 71
\bibitem[\protect\citeauthoryear{Quang~Huy and
  Kien~Phuc}{2025}]{quang2025insight}
\begin{botherref}
\oauthor{\bsnm{Quang~Huy}, \binits{P.}},
\oauthor{\bsnm{Kien~Phuc}, \binits{V.}}:
Insight into how legal and ethical considerations of artificial intelligence
  enhance the effectiveness of cyber forensic accounting.
Journal of Global Information Technology Management,
1--31
(2025)
\end{botherref}
\endbibitem

%%% 72
\bibitem[\protect\citeauthoryear{Wischmeyer}{2019}]{wischmeyer2019artificial}
\begin{bchapter}
\bauthor{\bsnm{Wischmeyer}, \binits{T.}}:
\bctitle{Artificial intelligence and transparency: opening the black box}.
In: \bbtitle{Regulating Artificial Intelligence},
pp. \bfpage{75}--\blpage{101}
(\byear{2019})
\end{bchapter}
\endbibitem

%%% 73
\bibitem[\protect\citeauthoryear{Djeffal}{2019}]{djeffal2019artificial}
\begin{bchapter}
\bauthor{\bsnm{Djeffal}, \binits{C.}}:
\bctitle{Artificial intelligence and public governance: normative guidelines
  for artificial intelligence in government and public administration}.
In: \bbtitle{Regulating Artificial Intelligence},
pp. \bfpage{277}--\blpage{293}
(\byear{2019})
\end{bchapter}
\endbibitem

%%% 74
\bibitem[\protect\citeauthoryear{Cath}{2018}]{cath2018governing}
\begin{barticle}
\bauthor{\bsnm{Cath}, \binits{C.}}:
\batitle{Governing artificial intelligence: ethical, legal and technical
  opportunities and challenges}.
\bjtitle{Philosophical Transactions of the Royal Society A: Mathematical,
  Physical and Engineering Sciences}
\bvolume{376}(\bissue{2133}),
\bfpage{20180080}
(\byear{2018})
\end{barticle}
\endbibitem

%%% 75
\bibitem[\protect\citeauthoryear{Baror et~al.}{2021}]{baror2021natural}
\begin{barticle}
\bauthor{\bsnm{Baror}, \binits{S.O.}},
\bauthor{\bsnm{Venter}, \binits{H.S.}},
\bauthor{\bsnm{Adeyemi}, \binits{R.}}:
\batitle{A natural human language framework for digital forensic readiness in
  the public cloud}.
\bjtitle{Australian Journal of Forensic Sciences}
\bvolume{53}(\bissue{5}),
\bfpage{566}--\blpage{591}
(\byear{2021})
\end{barticle}
\endbibitem

%%% 76
\bibitem[\protect\citeauthoryear{Jain}{2024}]{jain2024enhancing}
\begin{botherref}
\oauthor{\bsnm{Jain}, \binits{A.}}:
Enhancing forensic analysis of digital evidence using machine learning:
  Techniques, applications, and challenges.
International Journal of Innovative Research in Multidisciplinary Perspectives
  and Studies (IJIRMPS),
1--8
(2024)
\end{botherref}
\endbibitem

%%% 77
\bibitem[\protect\citeauthoryear{Wang and Zhang}{2024}]{wang2024group}
\begin{bchapter}
\bauthor{\bsnm{Wang}, \binits{Z.}},
\bauthor{\bsnm{Zhang}, \binits{W.}}:
\bctitle{Group fairness with individual and censorship constraints}.
In: \bbtitle{27th European Conference on Artificial Intelligence}
(\byear{2024})
\end{bchapter}
\endbibitem

%%% 78
\bibitem[\protect\citeauthoryear{Wang et~al.}{2024}]{wang2024individual}
\begin{bchapter}
\bauthor{\bsnm{Wang}, \binits{Z.}},
\bauthor{\bsnm{Ulloa}, \binits{D.}},
\bauthor{\bsnm{Yu}, \binits{T.}},
\bauthor{\bsnm{Rangaswami}, \binits{R.}},
\bauthor{\bsnm{Yap}, \binits{R.}},
\bauthor{\bsnm{Zhang}, \binits{W.}}:
\bctitle{Individual fairness with group constraints in graph neural networks}.
In: \bbtitle{27th European Conference on Artificial Intelligence}
(\byear{2024})
\end{bchapter}
\endbibitem

%%% 79
\bibitem[\protect\citeauthoryear{Yin et~al.}{2024}]{yin2024accessible}
\begin{bchapter}
\bauthor{\bsnm{Yin}, \binits{Z.}},
\bauthor{\bsnm{Agarwal}, \binits{S.}},
\bauthor{\bsnm{Kashif}, \binits{A.}},
\bauthor{\bsnm{Gonzalez}, \binits{M.}},
\bauthor{\bsnm{Wang}, \binits{Z.}},
\bauthor{\bsnm{Liu}, \binits{S.}},
\bauthor{\bsnm{Liu}, \binits{Z.}},
\bauthor{\bsnm{Wu}, \binits{Y.}},
\bauthor{\bsnm{Stockwell}, \binits{I.}},
\bauthor{\bsnm{Xu}, \binits{W.}}, \betal:
\bctitle{Accessible health screening using body fat estimation by image
  segmentation}.
In: \bbtitle{2024 IEEE International Conference on Data Mining Workshops
  (ICDMW)},
pp. \bfpage{405}--\blpage{414}
(\byear{2024})
\end{bchapter}
\endbibitem

%%% 80
\bibitem[\protect\citeauthoryear{Wang et~al.}{2025a}]{wang2025fg}
\begin{barticle}
\bauthor{\bsnm{Wang}, \binits{Z.}},
\bauthor{\bsnm{Yin}, \binits{Z.}},
\bauthor{\bsnm{Zhang}, \binits{Y.}},
\bauthor{\bsnm{Yang}, \binits{L.}},
\bauthor{\bsnm{Zhang}, \binits{T.}},
\bauthor{\bsnm{Pissinou}, \binits{N.}},
\bauthor{\bsnm{Cai}, \binits{Y.}},
\bauthor{\bsnm{Hu}, \binits{S.}},
\bauthor{\bsnm{Li}, \binits{Y.}},
\bauthor{\bsnm{Zhao}, \binits{L.}}, \betal:
\batitle{Fg-smote: Towards fair node classification with graph neural network}.
\bjtitle{ACM SIGKDD Explorations Newsletter}
\bvolume{26}(\bissue{2}),
\bfpage{99}--\blpage{108}
(\byear{2025})
\end{barticle}
\endbibitem

%%% 81
\bibitem[\protect\citeauthoryear{Wang et~al.}{2025b}]{wang2025graph}
\begin{barticle}
\bauthor{\bsnm{Wang}, \binits{Z.}},
\bauthor{\bsnm{Yin}, \binits{Z.}},
\bauthor{\bsnm{Liu}, \binits{F.}},
\bauthor{\bsnm{Liu}, \binits{Z.}},
\bauthor{\bsnm{Lisetti}, \binits{C.}},
\bauthor{\bsnm{Yu}, \binits{R.}},
\bauthor{\bsnm{Wang}, \binits{S.}},
\bauthor{\bsnm{Liu}, \binits{J.}},
\bauthor{\bsnm{Ganapati}, \binits{S.}},
\bauthor{\bsnm{Zhou}, \binits{S.}}, \betal:
\batitle{Graph fairness via authentic counterfactuals: Tackling structural and
  causal challenges}.
\bjtitle{ACM SIGKDD Explorations Newsletter}
\bvolume{26}(\bissue{2}),
\bfpage{89}--\blpage{98}
(\byear{2025})
\end{barticle}
\endbibitem

%%% 82
\bibitem[\protect\citeauthoryear{Wang et~al.}{2025c}]{wang2025fair}
\begin{bchapter}
\bauthor{\bsnm{Wang}, \binits{Z.}},
\bauthor{\bsnm{Chu}, \binits{Z.}},
\bauthor{\bsnm{Viet~Doan}, \binits{T.}},
\bauthor{\bsnm{Wang}, \binits{S.}},
\bauthor{\bsnm{Wu}, \binits{Y.}},
\bauthor{\bsnm{Palade}, \binits{V.}},
\bauthor{\bsnm{Zhang}, \binits{W.}}:
\bctitle{Fair graph u-net: A fair graph learning framework integrating group
  and individual awareness}.
In: \bbtitle{Proceedings of the AAAI Conference on Artificial Intelligence}
(\byear{2025})
\end{bchapter}
\endbibitem

%%% 83
\bibitem[\protect\citeauthoryear{Wang et~al.}{2025d}]{wang2025towards}
\begin{bchapter}
\bauthor{\bsnm{Wang}, \binits{Z.}},
\bauthor{\bsnm{Hoang}, \binits{N.}},
\bauthor{\bsnm{Zhang}, \binits{X.}},
\bauthor{\bsnm{Bello}, \binits{K.}},
\bauthor{\bsnm{Zhang}, \binits{X.}},
\bauthor{\bsnm{Iyengar}, \binits{S.S.}},
\bauthor{\bsnm{Zhang}, \binits{W.}}:
\bctitle{Towards fair graph learning without demographic information}.
In: \bbtitle{The 28th International Conference on Artificial Intelligence and
  Statistics}
(\byear{2025})
\end{bchapter}
\endbibitem

%%% 84
\bibitem[\protect\citeauthoryear{Zhang}{2024a}]{zhang2024fairness}
\begin{bchapter}
\bauthor{\bsnm{Zhang}, \binits{W.}}:
\bctitle{Fairness with censorship: Bridging the gap between fairness research
  and real-world deployment}.
In: \bbtitle{Proceedings of the AAAI Conference on Artificial Intelligence},
vol. \bseriesno{38},
pp. \bfpage{22685}--\blpage{22685}
(\byear{2024})
\end{bchapter}
\endbibitem

%%% 85
\bibitem[\protect\citeauthoryear{Zhang}{2024b}]{zhang2024inpractice}
\begin{botherref}
\oauthor{\bsnm{Zhang}, \binits{W.}}:
Ai fairness in practice: Paradigm, challenges, and prospects.
Ai Magazine
(2024)
\end{botherref}
\endbibitem

%%% 86
\bibitem[\protect\citeauthoryear{Zhang and Ntoutsi}{2019}]{zhang2019faht}
\begin{bchapter}
\bauthor{\bsnm{Zhang}, \binits{W.}},
\bauthor{\bsnm{Ntoutsi}, \binits{E.}}:
\bctitle{Faht: an adaptive fairness-aware decision tree classifier}.
In: \bbtitle{International Joint Conference on Artificial Intelligence
  (IJCAI)},
pp. \bfpage{1480}--\blpage{1486}
(\byear{2019})
\end{bchapter}
\endbibitem

%%% 87
\bibitem[\protect\citeauthoryear{Zhang and Weiss}{2021}]{zhang2021fair}
\begin{bchapter}
\bauthor{\bsnm{Zhang}, \binits{W.}},
\bauthor{\bsnm{Weiss}, \binits{J.}}:
\bctitle{Fair decision-making under uncertainty}.
In: \bbtitle{{2021 IEEE International Conference on Data Mining (ICDM)}}
(\byear{2021}).
\bcomment{IEEE}
\end{bchapter}
\endbibitem

%%% 88
\bibitem[\protect\citeauthoryear{Zhang and Weiss}{2022}]{zhang2022longitudinal}
\begin{bchapter}
\bauthor{\bsnm{Zhang}, \binits{W.}},
\bauthor{\bsnm{Weiss}, \binits{J.C.}}:
\bctitle{Longitudinal fairness with censorship}.
In: \bbtitle{Proceedings of the AAAI Conference on Artificial Intelligence},
vol. \bseriesno{36},
pp. \bfpage{12235}--\blpage{12243}
(\byear{2022})
\end{bchapter}
\endbibitem

%%% 89
\bibitem[\protect\citeauthoryear{Zhang et~al.}{2023}]{zhang2023censored}
\begin{bchapter}
\bauthor{\bsnm{Zhang}, \binits{W.}},
\bauthor{\bsnm{Hernandez-Boussard}, \binits{T.}},
\bauthor{\bsnm{Weiss}, \binits{J.}}:
\bctitle{Censored fairness through awareness}.
In: \bbtitle{Proceedings of the AAAI Conference on Artificial Intelligence},
vol. \bseriesno{37},
pp. \bfpage{14611}--\blpage{14619}
(\byear{2023})
\end{bchapter}
\endbibitem

%%% 90
\bibitem[\protect\citeauthoryear{Zhang et~al.}{2025}]{zhang2025fairness}
\begin{barticle}
\bauthor{\bsnm{Zhang}, \binits{W.}},
\bauthor{\bsnm{Zhou}, \binits{S.}},
\bauthor{\bsnm{Walsh}, \binits{T.}},
\bauthor{\bsnm{Weiss}, \binits{J.C.}}:
\batitle{Fairness amidst non-iid graph data: A literature review}.
\bjtitle{AI Magazine}
\bvolume{46}(\bissue{1}),
\bfpage{12212}
(\byear{2025})
\end{barticle}
\endbibitem

%%% 91
\bibitem[\protect\citeauthoryear{Zhang and Weiss}{2023}]{zhang2023fairness}
\begin{botherref}
\oauthor{\bsnm{Zhang}, \binits{W.}},
\oauthor{\bsnm{Weiss}, \binits{J.C.}}:
Fairness with censorship and group constraints.
Knowledge and Information Systems,
1--24
(2023)
\end{botherref}
\endbibitem

%%% 92
\bibitem[\protect\citeauthoryear{Zhang et~al.}{2021}]{zhang2021disentangled}
\begin{bchapter}
\bauthor{\bsnm{Zhang}, \binits{W.}},
\bauthor{\bsnm{Zhang}, \binits{L.}},
\bauthor{\bsnm{Pfoser}, \binits{D.}},
\bauthor{\bsnm{Zhao}, \binits{L.}}:
\bctitle{Disentangled dynamic graph deep generation}.
In: \bbtitle{Proceedings of the 2021 SIAM International Conference on Data
  Mining (SDM)},
pp. \bfpage{738}--\blpage{746}
(\byear{2021}).
\bcomment{SIAM}
\end{bchapter}
\endbibitem

%%% 93
\bibitem[\protect\citeauthoryear{Casey}{2011}]{casey2011digital}
\begin{bbook}
\bauthor{\bsnm{Casey}, \binits{E.}}:
\bbtitle{Digital Evidence and Computer Crime: Forensic Science, Computers, and
  the Internet},
(\byear{2011})
\end{bbook}
\endbibitem

%%% 94
\bibitem[\protect\citeauthoryear{Walker}{2001}]{walker2001digital}
\begin{barticle}
\bauthor{\bsnm{Walker}, \binits{C.}}:
\batitle{Digital evidence and computer crime: Forensic science, computers and
  the internet}.
\bjtitle{Crime Prevention and Community Safety}
\bvolume{3},
\bfpage{87}--\blpage{88}
(\byear{2001})
\end{barticle}
\endbibitem

%%% 95
\bibitem[\protect\citeauthoryear{Sharevski}{2015}]{sharevski2015rules}
\begin{barticle}
\bauthor{\bsnm{Sharevski}, \binits{F.}}:
\batitle{Rules of professional responsibility in digital forensics: A
  comparative analysis}.
\bjtitle{Journal of Digital Forensics, Security and Law}
\bvolume{10}(\bissue{2}),
\bfpage{3}
(\byear{2015})
\end{barticle}
\endbibitem

%%% 96
\bibitem[\protect\citeauthoryear{Ademu et~al.}{2011}]{ademu2011new}
\begin{botherref}
\oauthor{\bsnm{Ademu}, \binits{I.O.}},
\oauthor{\bsnm{Imafidon}, \binits{C.O.}},
\oauthor{\bsnm{Preston}, \binits{D.S.}}:
A new approach of digital forensic model for digital forensic investigation.
International Journal of Advanced Computer Science and Applications
\textbf{2}(12)
(2011)
\end{botherref}
\endbibitem

%%% 97
\bibitem[\protect\citeauthoryear{Quick and Choo}{2014a}]{quick2014data}
\begin{botherref}
\oauthor{\bsnm{Quick}, \binits{D.}},
\oauthor{\bsnm{Choo}, \binits{K.-K.R.}}:
Data reduction and data mining framework for digital forensic evidence:
  storage, intelligence, review and archive.
Trends and Issues in Crime and Criminal Justice
(480),
1--11
(2014)
\end{botherref}
\endbibitem

%%% 98
\bibitem[\protect\citeauthoryear{Quick and Choo}{2014b}]{quick2014impacts}
\begin{barticle}
\bauthor{\bsnm{Quick}, \binits{D.}},
\bauthor{\bsnm{Choo}, \binits{K.-K.R.}}:
\batitle{Impacts of increasing volume of digital forensic data: A survey and
  future research challenges}.
\bjtitle{Digital Investigation}
\bvolume{11}(\bissue{4}),
\bfpage{273}--\blpage{294}
(\byear{2014})
\end{barticle}
\endbibitem

%%% 99
\bibitem[\protect\citeauthoryear{Lillis et~al.}{2016}]{lillis2016current}
\begin{botherref}
\oauthor{\bsnm{Lillis}, \binits{D.}},
\oauthor{\bsnm{Becker}, \binits{B.}},
\oauthor{\bsnm{O'Sullivan}, \binits{T.}},
\oauthor{\bsnm{Scanlon}, \binits{M.}}:
Current challenges and future research areas for digital forensic
  investigation.
arXiv preprint arXiv:1604.03850
(2016)
\end{botherref}
\endbibitem

%%% 100
\bibitem[\protect\citeauthoryear{Vincze}{2016}]{vincze2016challenges}
\begin{barticle}
\bauthor{\bsnm{Vincze}, \binits{E.A.}}:
\batitle{Challenges in digital forensics}.
\bjtitle{Police Practice and Research}
\bvolume{17}(\bissue{2}),
\bfpage{183}--\blpage{194}
(\byear{2016})
\end{barticle}
\endbibitem

%%% 101
\bibitem[\protect\citeauthoryear{Rowlingson et~al.}{2004}]{rowlingson2004ten}
\begin{barticle}
\bauthor{\bsnm{Rowlingson}, \binits{R.}}, \betal:
\batitle{A ten step process for forensic readiness}.
\bjtitle{International Journal of Digital Evidence}
\bvolume{2}(\bissue{3}),
\bfpage{1}--\blpage{28}
(\byear{2004})
\end{barticle}
\endbibitem

%%% 102
\bibitem[\protect\citeauthoryear{Amato et~al.}{2017}]{amato2017correlation}
\begin{bchapter}
\bauthor{\bsnm{Amato}, \binits{F.}},
\bauthor{\bsnm{Cozzolino}, \binits{G.}},
\bauthor{\bsnm{Mazzeo}, \binits{A.}},
\bauthor{\bsnm{Mazzocca}, \binits{N.}}:
\bctitle{Correlation of digital evidences in forensic investigation through
  semantic technologies}.
In: \bbtitle{2017 31st International Conference on Advanced Information
  Networking and Applications Workshops (WAINA)},
pp. \bfpage{668}--\blpage{673}
(\byear{2017}).
\bcomment{IEEE}
\end{bchapter}
\endbibitem

%%% 103
\bibitem[\protect\citeauthoryear{Horsman}{2024}]{horsman2024importance}
\begin{barticle}
\bauthor{\bsnm{Horsman}, \binits{G.}}:
\batitle{The importance of digital evidence strategies}.
\bjtitle{Wiley Interdisciplinary Reviews: Forensic Science}
\bvolume{6}(\bissue{1}),
\bfpage{1507}
(\byear{2024})
\end{barticle}
\endbibitem

%%% 104
\bibitem[\protect\citeauthoryear{Ferrag et~al.}{2024}]{ferrag2024generative}
\begin{botherref}
\oauthor{\bsnm{Ferrag}, \binits{M.A.}},
\oauthor{\bsnm{Alwahedi}, \binits{F.}},
\oauthor{\bsnm{Battah}, \binits{A.}},
\oauthor{\bsnm{Cherif}, \binits{B.}},
\oauthor{\bsnm{Mechri}, \binits{A.}},
\oauthor{\bsnm{Tihanyi}, \binits{N.}}:
Generative ai and large language models for cyber security: All insights you
  need.
Available at SSRN 4853709
(2024)
\end{botherref}
\endbibitem

%%% 105
\bibitem[\protect\citeauthoryear{Kucharavy
  et~al.}{2023}]{kucharavy2023fundamentals}
\begin{botherref}
\oauthor{\bsnm{Kucharavy}, \binits{A.}},
\oauthor{\bsnm{Schillaci}, \binits{Z.}},
\oauthor{\bsnm{Mar{\'e}chal}, \binits{L.}},
\oauthor{\bsnm{W{\"u}rsch}, \binits{M.}},
\oauthor{\bsnm{Dolamic}, \binits{L.}},
\oauthor{\bsnm{Sabonnadiere}, \binits{R.}},
\oauthor{\bsnm{David}, \binits{D.P.}},
\oauthor{\bsnm{Mermoud}, \binits{A.}},
\oauthor{\bsnm{Lenders}, \binits{V.}}:
Fundamentals of generative large language models and perspectives in
  cyber-defense.
arXiv preprint arXiv:2303.12132
(2023)
\end{botherref}
\endbibitem

%%% 106
\bibitem[\protect\citeauthoryear{Zhao et~al.}{2023}]{zhao2023survey}
\begin{botherref}
\oauthor{\bsnm{Zhao}, \binits{W.X.}},
\oauthor{\bsnm{Zhou}, \binits{K.}},
\oauthor{\bsnm{Li}, \binits{J.}},
\oauthor{\bsnm{Tang}, \binits{T.}},
\oauthor{\bsnm{Wang}, \binits{X.}},
\oauthor{\bsnm{Hou}, \binits{Y.}},
\oauthor{\bsnm{Min}, \binits{Y.}},
\oauthor{\bsnm{Zhang}, \binits{B.}},
\oauthor{\bsnm{Zhang}, \binits{J.}},
\oauthor{\bsnm{Dong}, \binits{Z.}}, et al.:
A survey of large language models.
arXiv preprint arXiv:2303.18223
\textbf{1}(2)
(2023)
\end{botherref}
\endbibitem

%%% 107
\bibitem[\protect\citeauthoryear{Chang et~al.}{2024}]{chang2024survey}
\begin{barticle}
\bauthor{\bsnm{Chang}, \binits{Y.}},
\bauthor{\bsnm{Wang}, \binits{X.}},
\bauthor{\bsnm{Wang}, \binits{J.}},
\bauthor{\bsnm{Wu}, \binits{Y.}},
\bauthor{\bsnm{Yang}, \binits{L.}},
\bauthor{\bsnm{Zhu}, \binits{K.}},
\bauthor{\bsnm{Chen}, \binits{H.}},
\bauthor{\bsnm{Yi}, \binits{X.}},
\bauthor{\bsnm{Wang}, \binits{C.}},
\bauthor{\bsnm{Wang}, \binits{Y.}}, \betal:
\batitle{A survey on evaluation of large language models}.
\bjtitle{ACM transactions on intelligent systems and technology}
\bvolume{15}(\bissue{3}),
\bfpage{1}--\blpage{45}
(\byear{2024})
\end{barticle}
\endbibitem

%%% 108
\bibitem[\protect\citeauthoryear{Naveed et~al.}{2023}]{naveed2023comprehensive}
\begin{botherref}
\oauthor{\bsnm{Naveed}, \binits{H.}},
\oauthor{\bsnm{Khan}, \binits{A.U.}},
\oauthor{\bsnm{Qiu}, \binits{S.}},
\oauthor{\bsnm{Saqib}, \binits{M.}},
\oauthor{\bsnm{Anwar}, \binits{S.}},
\oauthor{\bsnm{Usman}, \binits{M.}},
\oauthor{\bsnm{Akhtar}, \binits{N.}},
\oauthor{\bsnm{Barnes}, \binits{N.}},
\oauthor{\bsnm{Mian}, \binits{A.}}:
A comprehensive overview of large language models.
arXiv preprint arXiv:2307.06435
(2023)
\end{botherref}
\endbibitem

%%% 109
\bibitem[\protect\citeauthoryear{Shanahan}{2024}]{shanahan2024talking}
\begin{barticle}
\bauthor{\bsnm{Shanahan}, \binits{M.}}:
\batitle{Talking about large language models}.
\bjtitle{Communications of the ACM}
\bvolume{67}(\bissue{2}),
\bfpage{68}--\blpage{79}
(\byear{2024})
\end{barticle}
\endbibitem

%%% 110
\bibitem[\protect\citeauthoryear{Zhao et~al.}{2024}]{zhao2024explainability}
\begin{barticle}
\bauthor{\bsnm{Zhao}, \binits{H.}},
\bauthor{\bsnm{Chen}, \binits{H.}},
\bauthor{\bsnm{Yang}, \binits{F.}},
\bauthor{\bsnm{Liu}, \binits{N.}},
\bauthor{\bsnm{Deng}, \binits{H.}},
\bauthor{\bsnm{Cai}, \binits{H.}},
\bauthor{\bsnm{Wang}, \binits{S.}},
\bauthor{\bsnm{Yin}, \binits{D.}},
\bauthor{\bsnm{Du}, \binits{M.}}:
\batitle{Explainability for large language models: A survey}.
\bjtitle{ACM Transactions on Intelligent Systems and Technology}
\bvolume{15}(\bissue{2}),
\bfpage{1}--\blpage{38}
(\byear{2024})
\end{barticle}
\endbibitem

%%% 111
\bibitem[\protect\citeauthoryear{Chen et~al.}{2021}]{chen2021evaluating}
\begin{botherref}
\oauthor{\bsnm{Chen}, \binits{M.}},
\oauthor{\bsnm{Tworek}, \binits{J.}},
\oauthor{\bsnm{Jun}, \binits{H.}},
\oauthor{\bsnm{Yuan}, \binits{Q.}},
\oauthor{\bsnm{Pinto}, \binits{H.P.D.O.}},
\oauthor{\bsnm{Kaplan}, \binits{J.}},
\oauthor{\bsnm{Edwards}, \binits{H.}},
\oauthor{\bsnm{Burda}, \binits{Y.}},
\oauthor{\bsnm{Joseph}, \binits{N.}},
\oauthor{\bsnm{Brockman}, \binits{G.}}, et al.:
Evaluating large language models trained on code.
arXiv preprint arXiv:2107.03374
(2021)
\end{botherref}
\endbibitem

%%% 112
\bibitem[\protect\citeauthoryear{Zhou et~al.}{}]{zhoullm}
\begin{botherref}
\oauthor{\bsnm{Zhou}, \binits{H.}},
\oauthor{\bsnm{Xu}, \binits{W.}},
\oauthor{\bsnm{Dehlinger}, \binits{J.}},
\oauthor{\bsnm{Chakraborty}, \binits{S.}},
\oauthor{\bsnm{Deng}, \binits{L.}}:
An llm-driven approach to gain cybercrime insights with evidence networks
\end{botherref}
\endbibitem

%%% 113
\bibitem[\protect\citeauthoryear{Chernyshev
  et~al.}{2023}]{chernyshev2023towards}
\begin{bchapter}
\bauthor{\bsnm{Chernyshev}, \binits{M.}},
\bauthor{\bsnm{Baig}, \binits{Z.}},
\bauthor{\bsnm{Doss}, \binits{R.R.M.}}:
\bctitle{Towards large language model (llm) forensics using llm-based
  invocation log analysis}.
In: \bbtitle{Proceedings of the 1st ACM Workshop on Large AI Systems and Models
  with Privacy and Safety Analysis},
pp. \bfpage{89}--\blpage{96}
(\byear{2023})
\end{bchapter}
\endbibitem

%%% 114
\bibitem[\protect\citeauthoryear{Walker et~al.}{2024}]{walker2024forensic}
\begin{bchapter}
\bauthor{\bsnm{Walker}, \binits{C.}},
\bauthor{\bsnm{Gharaibeh}, \binits{T.}},
\bauthor{\bsnm{Alsmadi}, \binits{R.}},
\bauthor{\bsnm{Hall}, \binits{C.}},
\bauthor{\bsnm{Baggili}, \binits{I.}}:
\bctitle{Forensic analysis of artifacts from microsoft's multi-agent llm
  platform autogen}.
In: \bbtitle{Proceedings of the 19th International Conference on Availability,
  Reliability and Security},
pp. \bfpage{1}--\blpage{9}
(\byear{2024})
\end{bchapter}
\endbibitem

%%% 115
\bibitem[\protect\citeauthoryear{Sharma et~al.}{2025}]{sharma2025forensicllm}
\begin{barticle}
\bauthor{\bsnm{Sharma}, \binits{B.}},
\bauthor{\bsnm{Ghawaly}, \binits{J.}},
\bauthor{\bsnm{McCleary}, \binits{K.}},
\bauthor{\bsnm{Webb}, \binits{A.M.}},
\bauthor{\bsnm{Baggili}, \binits{I.}}:
\batitle{Forensicllm: A local large language model for digital forensics}.
\bjtitle{Forensic Science International: Digital Investigation}
\bvolume{52},
\bfpage{301872}
(\byear{2025})
\end{barticle}
\endbibitem

%%% 116
\bibitem[\protect\citeauthoryear{Saxena et~al.}{2023}]{saxena2023missed}
\begin{bchapter}
\bauthor{\bsnm{Saxena}, \binits{N.A.}},
\bauthor{\bsnm{Zhang}, \binits{W.}},
\bauthor{\bsnm{Shahabi}, \binits{C.}}:
\bctitle{Missed opportunities in fair ai}.
In: \bbtitle{Proceedings of the 2023 SIAM International Conference on Data
  Mining (SDM)},
pp. \bfpage{961}--\blpage{964}
(\byear{2023}).
\bcomment{SIAM}
\end{bchapter}
\endbibitem

\end{thebibliography}

\end{document}